\documentclass[amsmath,amssymb,superscriptaddress]{revtex4-2}
\usepackage{url}
\usepackage{setspace} 
\usepackage{lmodern}
\usepackage{graphicx}
\usepackage{gensymb}

\usepackage{amsmath}
\usepackage{newtxmath}
\usepackage[colorlinks=true, citecolor=blue, linkcolor=blue, urlcolor=red]{hyperref}

\usepackage{natbib}
\begin{document}

\title{Diffusioosmotic dispersion}

\author{Jian Teng}
\affiliation{Center for Fluid Mechanics, Brown University, Providence, Rhode Island 02912, USA}
\author{Bhargav Rallabandi}
\affiliation{Department of Mechanical Engineering, University of California, Riverside, CA 92521, USA}
\author{Jesse T. Ault}
\email{jesse\_ault@brown.edu}
\affiliation{Center for Fluid Mechanics, Brown University, Providence, Rhode Island 02912, USA}
\begin{abstract}

Solute-surface interactions have garnered considerable interest in recent years as a novel control mechanism for driving unique fluid dynamics and particle transport with potential applications in fields such as biomedicine, the development of microfluidic devices, and enhanced oil recovery. In this study, we will discuss dispersion induced by the diffusioosmotic motion near a charged wall in the presence of a solute concentration gradient. Here, we introduce a plug of salt with a Gaussian distribution at the center of a channel with no background flow. As the solute diffuses, the concentration gradient drives a diffusioosmotic slip flow at the walls, which results in a recirculating flow in the channel; this, in turn, drives an advective flux of the solute concentration. This effect leads to cross-stream diffusion of the solute, altering the effective diffusivity of the solute as it diffuses along the channel. We derive theoretical predictions for the solute dynamics using a multiple-timescale analysis to quantify the dispersion driven by the solute-surface interactions. Furthermore, we derive a cross-sectionally averaged concentration equation with an effective diffusivity analogous to that from Taylor dispersion. In addition, we use numerical simulations to validate our theoretical predictions.
\end{abstract}

\maketitle

\section{Introduction}
\label{sec:intro}
In fluid dynamics, dispersion is typically used to denote transport of a species from high to low concentrations due to non-uniform flow conditions. This is in contrast to diffusion, which denotes the similar transport of a species from high to low concentrations but due to Brownian motion. Since relatively few flow systems actually transport flow in a plug-like way, dispersion is a nearly ubiquitous transport mechanism in systems including microfluidic devices, filtration systems, chemical reactor systems, medical devices, lab-on-a-chip systems, and many others. A classical and simple example of dispersion is that which has come to be known as Taylor dispersion, which was first studied by \citet{taylor1953dispersion} and later generalized by \citet{aris1956dispersion}. Taylor dispersion describes the enhanced diffusivity that a diffusing species experiences in the presence of a shear flow, such as the diffusion of a solute concentration field in a pipe or channel flow. In both of these cases, the advection-diffusion equation governing the solute transport can be averaged over the cross-section to yield a 1D model for the depth-averaged concentration, which experiences an enhanced effective diffusivity that is a function of the Pecl\'{e}t number governing the transport. A simple physical understanding of this Taylor dispersion can be achieved by supposing we have a step initial condition in the concentration $c$ of some solute species. For example, suppose we have a Poiseuille flow in a pipe, and we suddenly add solute to the system such that $c(x<0)=c_1$ and $c(x>=0)=c_2$, where $-\infty<x<\infty$ is the axial coordinate of our infinite pipe. Then, in the limit of no background flow, the interface between the two solute concentrations will smear out by diffusion in a purely 1D process. However, as the relative importance of fluid advection to solute diffusion (i.e., the Pecl\'{e}t number) increases, shear flows in the system distort the interface between the two solute concentrations as they diffuse, introducing cross-stream diffusion and enhancing the rate of axial diffusion of the cross-sectionally averaged concentration. For more rigorous background into the Taylor dispersion phenomenon, see (1) \citet{barton1983method} who extended the methods and results of \citet{aris1956dispersion} to consider all times rather than just the asymptotic behavior, (2) \citet{frankel1989foundations} who developed a generalized Taylor dispersion theory to greatly extend the ideas of Taylor and Aris to whole classes of other problems such as porous media flows or even sedimenting particles, and (3) \citet{brenner1993macrotransport} who provide a comprehensive overview of the theory of macrotransport processes.

Many other studies and applications have built upon the theoretical work on Taylor dispersion by Taylor and Aris. Here we just briefly mention a few examples. \citet{stone1999dispersion} extended the theories of Taylor dispersion to consider laminar flows with velocity variations in the streamwise direction. \citet{aminian2016boundaries} studied the role of the channel boundary shape and aspect ratio on the dispersion as a means to control the delivery of chemicals in microfluidics. \citet{salmon2020buoyancy} studied the solute dispersion induced by buoyancy-driven flow and developed analytical methods analogous to Taylor dispersion. \citet{chu2019dispersion} added the ideas of oscillatory pressure-driven flow in a parallel-plate channel flow as well as patterned slip walls and investigated the role of both of these effects on the dispersion process. Finally, \citet{chu2021macrotransport} coupled the Taylor dispersion in a pipe flow to the transport of a second species consisting of particles or bacteria via diffusiophoresis, which was also explored by \citet{migacz2022diffusiophoresis}. This list is by no means exhaustive, but one unifying theme that is common to many of the studies related to dispersion is the role of imposed pressure gradients or moving boundaries to drive the shear flows that cause the dispersion. Typically, the transport of the solute is passive in the sense that it is not expected to couple to and alter the background fluid dynamics. However, there are many scenarios where the transport of solute is fully coupled to the fluid dynamics, which is the focus of this paper. In particular, we consider the case where there is no background pressure-driven flow and no moving boundaries, but where the solute interacts with the boundaries of the system to drive fluid flow via diffusioosmosis which is the spontaneous motion of fluid near a surface in the presence of a solute concentration gradient.

The physical origin of diffusioosmosis was discovered by \citet{derjaguin1961diffusiophoresis}, where Derjaguin and his coworkers showed experimentally that a local solute concentration gradient near a boundary could induce a slip-flow boundary condition over the surface. Since then, significant theoretical progress has been made towards understanding and modeling this effect, and in the context of dispersion, a variety of studies have demonstrated the potential for diffusioosmosis to alter the transport of suspended species in confined geometries \citep{rasmussen2020size, ault2019characterization, shin2017accumulation, shin2016size, keh2005diffusioosmosis, kar2015enhanced,chakra2023continuous}. Diffusioosmosis is also closely related to the analogous phenomenon of diffusiophoresis. Whereas diffusioosmosis refers to the motion of fluid next to a surface in the presence of a chemical concentration gradient, diffusiophoresis refers to the reciprocal motion of suspended particles in a concentration gradient, which results due to the slip flow at the surface. Both diffusioosmosis and diffusiophoresis can have contributions due to chemi-osmosis/-phoresis that arise from the osmotic pressure gradient over the surface and electro-osmosis/-phoresis that arise in the case of electrolyte solutes with mismatched anion and cation diffusivities \citep{anderson1984diffusiophoresis}. Coupled diffusioosmosis and diffusiophoresis have been the subject of a variety of recent studies, especially concerning the coupled transport of solutes and colloidal particles in confined geometries where convective transport is difficult to achieve \citep{shin2018cleaning, kar2015enhanced, shin2017low, singh2020reversible, williams2020diffusioosmotic, alessio2021diffusiophoresis,chakra2023continuous}.

Such coupled motions have also been the focus of studies in a variety of other natural and engineering settings, such as in underground porous media flows \citep{park2021microfluidic}, water filtration systems \citep{florea2014long, shin2017accumulation, lee2018diffusiophoretic, bone2020advanced}, microfluidic devices \citep{palacci2010colloidal, shin2017membraneless}, fabric cleaning systems \citep{shin2018cleaning}, particle delivery methods \citep{ault2017diffusiophoresis}, energy storage applications \citep{gupta2020charging, gupta2020ionic}, and many others. Typically, in such studies the role of diffusioosmosis has been a secondary effect, or neglected entirely, although a collection of studies on the motion of solutes and particles in narrow pores has considered both effects \citep{shin2017low, singh2020reversible, ault2017diffusiophoresis, williams2020diffusioosmotic, alessio2021diffusiophoresis, alessio2022diffusioosmosis}. In such systems, diffusioosmosis can play an essential role on the transport of both the solutes, the fluid, and any suspended particles \citep{shim2022diffusiophoresis}. In the present study, we investigate the dispersion of solute in a channel where the only fluid motion is driven by diffusioosmosis at the channel walls. That is, the diffusion of the initial solute concentration results in a local concentration gradient at the walls that in turn drives a fluid recirculation via diffusioosmosis. The resulting shear flows then in turn alter the transport of the solute concentration by a mechanism analogous to that of Taylor dispersion.

As mentioned, several studies have already considered the coupling between Taylor dispersion and diffusiophoresis. Specifically, \citet{chu2021macrotransport} studied the diffusiophoretic dynamics of charged colloidal particles or bacteria in a solute concentration field that was experiencing Taylor dispersion. This work considered a one-way coupling where the fluid flow drives Taylor dispersion of the solute field and the particle concentration field, and the particles receive an additional contribution to their motion from diffusiophoresis via the solute field. They developed theoretical and numerical solutions in the long-time regime following an approach similar to that of Taylor and Aris. More recently, \citet{migacz2022diffusiophoresis} built upon this work by developing solutions that are valid for both the early- and long-time regimes. However, neither of these studies considered the diffusioosmosis at the channel walls. One such study that did consider these effects was that of \citet{alessio2022diffusioosmosis}, who considered the diffusioosmotic slip boundary condition and derived a multi-dimensional effective dispersion equation for solute transport into a dead-end pore similar to that of Taylor. In contrast to the previous studies, while the velocity profile in Alessio's work is still approximately parabolic (as in classical Taylor dispersion), the magnitude is a function of position and time in the channel as the solute concentration evolves. We will find a similar behavior in our system.

To the best of our knowledge, no theoretical solutions have previously been developed to describe the diffusioosmosis-driven analog of Taylor dispersion. In this study, we take motivations from the works of \citet{alessio2022diffusioosmosis}, \citet{chu2021macrotransport}, and \citet{migacz2022diffusiophoresis} and develop analytical solutions for the diffusioosmosis-driven dispersion of a plug of solute in a channel. We consider a plug of solute that is initially normally distributed in a Gaussian peak at the center of the channel (see figure \ref{fig:geo}). As the solute diffuses, the local concentration gradient at the wall drives an effective slip boundary condition via diffusioosmosis that is dependent on the charge of the surface. This slip at the walls drives a recirculating flow in the channel. The recirculation contributes to the advection of the solute transport and introduces cross-stream diffusion, which alters the effective diffusivity of the solute along the channel. In the modeling process, we use a perturbation method to derive analytical solutions to the coupled fluid and solute dynamics. The theoretical analysis is performed for transport in a long, narrow 2D channel using a 2D Cartesian coordinate system and in a long, narrow cylindrical pipe using a 2D axisymmetric cylindrical coordinate system. In $\S$ \ref{sec:theory}, we introduce the governing equations and boundary conditions for both systems. We then apply a perturbation method along with a multiple timescale analysis to theoretically solve for the fluid and solute dynamics. In $\S$ \ref{sec:ave}, we derive the effective diffusivity of the cross-sectionally averaged solute concentration analogously to that of Taylor dispersion. In $\S$ \ref{sec:numerical}, we perform numerical simulations to solve for the fluid and solute dynamics and show good agreement with the theoretical predictions. In $\S$ \ref{sec:result}, we analyze the dispersion behavior for various conditions and in different time regimes.

\section{Modeling diffusioosmotic dispersion in a long, narrow channel}\label{sec:theory}
\begin{figure}
    \centering
    \includegraphics[width=\textwidth]{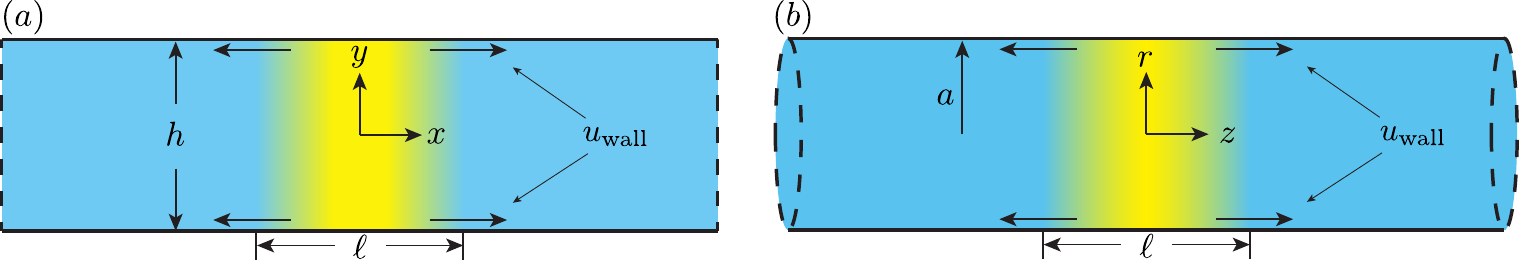}
    \caption{Problem setup. We consider an initially Gaussian distributed plug of salt with characteristic length $\ell$ in both (a) 2D Cartesian coordinates and (b) axisymmetric cylindrical coordinates. Both channels are infinite in the axial direction, and $u_\textrm{wall}$ represents the slip velocity at the wall induced by diffusioosmosis.}
    \label{fig:geo}
\end{figure}

In this section, we model the coupled fluid and solute transport for a diffusing plug of solute in a channel in the presence of diffusioosmosis-driven recirculation. We consider two configurations corresponding to planar and cylindrical channels and describe the flows in these configurations using Cartesian and cylindrical coordinates, respectively. The channel configurations for both systems are shown in figure \ref{fig:geo}. Initially, we introduce a plug of solute with a Gaussian distribution centered around the origin. In cases when the solute molecules/ions do not interact with the channel walls, the dynamics of the solute transport are governed by simple Brownian diffusion. Here, however, we consider the case where the channel walls have a non-zero surface charge and solute-surface interactions cannot be neglected. The local solute concentration gradient at the channel walls will induce a diffusioosmotic slip velocity boundary condition, which will in turn drive a recirculation in the channel as the solute diffuses. The magnitude of this slip velocity boundary condition is given by $u_\textrm{wall} = \Gamma_w \nabla_\parallel \ln c$, where $u_\textrm{wall}$ is the velocity at the wall in the direction parallel to the wall, and the gradient is taken parallel to the surface. Here, $\Gamma_w$ is the diffusioosmotic mobility coefficient, which is a function of the surface charge of the channel walls, and $c$ is the solute ion concentration.

\subsection{Governing equations}\label{sec:govn}
The fluid and solute dynamics in the system are governed by the coupled continuity and incompressible Navier-Stokes equations and an advection-diffusion equation for the solute transport. Here, the fluid pressure, density, viscosity, and velocity are given respectively by $p$, $\rho$, $\mu$, and $\boldsymbol{u} = (u,v)~\textrm{or}~(u_r,u_z)$. The solute concentration and diffusivity are given by $c(x,y,t)$ and $D_s$, respectively. Here, we consider $Re = \frac{\rho U h}{\mu} \ll 1$, where $U$ is some characteristic flow velocity in the axial direction, we neglect the influence of gravity, and we treat the fluid dynamics as quasi-steady. The dimensional form of the governing equations is given by
\begin{subequations} \label{dim_governing}
\begin{gather}
\nabla^*\cdot\boldsymbol{u}^*=0,\\
-\nabla^*p^*+\mu\nabla^{*2}\boldsymbol{u}^*=\boldsymbol{0},\\
\frac{\partial c^*}{\partial t^*}+ \nabla^*\cdot(\boldsymbol{u^*}c^*)=D_s\nabla^{*2}c^*,
\end{gather}
\end{subequations}
where asterisks denote dimensional quantities. The 2D axisymmetric cylindrical and the 2D Cartesian cases can be solved similarly and share similar boundary conditions. For the sake of simplicity, we only show the derivation for the 2D channel flow case in the main text and refer the reader to Appendix \ref{Appendix:B} for the derivation for the pipe flow case. The long, narrow channel has a height of $h$ in the Cartesian coordinate system. The channel is infinitely long, and the initial Gaussian solute distribution has a characteristic width of $\ell$, and we will seek solutions in the limit that $h\ll\ell$. To begin, we nondimensionalize the system of equations as follows,
\begin{equation}\label{eq:2D_nondim_coeff}
\begin{split}
    & x = \frac{x^*}{\ell},\; y = \frac{y^*}{h},\; u = \frac{u^*}{U},\; v = \frac{v^*\ell}{Uh},\; p = \frac{p^* h^2}{\mu U \ell},\; \epsilon = \frac{h}{\ell},\; U = \frac{D_s}{\ell},\; t = \frac{t^*}{\ell^2/D_s},
\end{split}
\end{equation}
where $U=D_s/\ell$ is the characteristic speed of solute diffusion along the channel, and $\ell^2/D_s$ is the characteristic time of diffusion along the channel. With these scalings, we can rewrite the governing equations (\ref{dim_governing}) in nondimensional form as,
\begin{subequations}\label{eq:non-2dim}
\begin{gather}
    0 = \frac{\partial u}{\partial x} + \frac{\partial v}{\partial y},\\
    0 = -\frac{\partial p}{\partial x} + \epsilon^2 \frac{\partial^2 u}{\partial x^2} + \frac{\partial^2 u}{\partial y^2},\\
    0 = -\frac{\partial p}{\partial y} + \epsilon^4 \frac{\partial^2 v}{\partial x^2} + \epsilon^2 \frac{\partial^2 v}{\partial y^2},\\
    \epsilon^2\frac{\partial c}{\partial t}+\epsilon^2 u \frac{\partial c}{\partial x} + \epsilon^2 v \frac{\partial c}{\partial y}=\epsilon^2 \frac{\partial^2 c}{\partial x^2} + \frac{\partial^2 c}{\partial y^2}.
\end{gather}
\end{subequations}

The solution to the governing equations (\ref{eq:non-2dim}) is subject to boundary conditions on the fluid and solute. These boundary conditions can be summarized by:
\begin{equation}
    \text{Quiescent far-field condition: }p=0,~~u=0,~~c=0~~\text{at}~~x=\pm\infty,
\end{equation}
\begin{equation} \label{bc:v}
    \text{No fluid penetration at the walls: } v = 0~\text{at}~y=\pm\frac{1}{2},
\end{equation}
\begin{equation} \label{c_bc}
    \text{No-flux conditions at the channel walls: }\frac{\partial c}{\partial y}=0~\text{at}~y=\pm\frac{1}{2},
\end{equation}
\begin{equation}\label{slip_condition}
\text{Diffusioosmotic wall slip boundary condition: }u = \frac{\Gamma_w}{D_s}\frac{\partial \ln c}{\partial x}~\text{at}~y=\pm\frac{1}{2}.
\end{equation}
The slip boundary condition given by (\ref{slip_condition}) is induced by diffusioosmosis, which drives the flow inside the channel. In this study, we assume a constant zeta potential and diffusioosmotic mobility coefficient $\Gamma_w$. This assumption is needed in order to achieve a final analytical solution, and is a reasonable approximation under many scenarios as discussed by several recent previous works (see, e.g., \citet{ault2017diffusiophoresis}, \citet{migacz2022diffusiophoresis}, \citet{Lee2022}, \citet{gupta2020diffusiophoresis}, and \citet{alessio2022diffusioosmosis}). We further focus our attention on cases where the initial condition is already spread out relative to the channel width, such that $\epsilon \ll 1$, in which case the lubrication approximation can be used to simplify the governing equations.

\subsection{Leading-order fluid dynamics}\label{sec:vel}
In our original nondimensionalization above, we chose a characteristic timescale that is the characteristic time for solute diffusion along the channel $\ell^2/D_s$. This corresponds to the slow dynamics of the system. The other important timescale in the system is the characteristic time for solute diffusion across the channel, $h^2/D_s$. This corresponds to the fast dynamics of the system, and the two timescales are separated by a factor of $\epsilon^2 = h^2/\ell^2 \ll 1$. Here, in order to develop a solution that is uniformly valid across both the early and late dynamics, we use an approach similar to that of \citet{migacz2022diffusiophoresis}.

Following this approach, we introduce a multiple timescale analysis \citep{bender1999advanced} in which we introduce a fast time variable $T=t/\epsilon^2$. That is, $T=O(1)$ over dimensional times $\sim h^2/D_s$ corresponding to $t=O(\epsilon^2)$, whereas $t=O(1)$ on dimensional times $\sim \ell^2/D_s$. Thus, we can map any time-dependent quantity as
\begin{equation}
    f(t)\mapsto f(t,T)\Rightarrow\frac{\partial f}{\partial t}\mapsto \frac{\partial f}{\partial t}+\frac{\partial T}{\partial t}\frac{\partial f}{\partial T} = \frac{\partial f}{\partial t}+\epsilon^{-2}\frac{\partial f}{\partial T}.
\end{equation}
With this mapping, the advection-diffusion equation (\ref{eq:non-2dim}$d$), becomes
\begin{equation}\label{adv-diff}
    \epsilon^2\frac{\partial c}{\partial t}+\frac{\partial c}{\partial T}+\epsilon^2 u \frac{\partial c}{\partial x} + \epsilon^2 v \frac{\partial c}{\partial y}=\epsilon^2 \frac{\partial^2 c}{\partial x^2} + \frac{\partial^2 c}{\partial y^2}.
\end{equation}
We seek an analytical solution of the governing equations as perturbation expansions in the small parameter $\epsilon^2$ in the limit of $Re \ll 1$. We seek solutions of the form
\begin{subequations}\label{perturbation}
\begin{align}
c(x,y,t,T) &= c_0(x,y,t,T) + \epsilon^2 c_1(x,y,t,T) + \epsilon^4 c_2(x,y,t,T) + \hdots,\\
p(x,y,t,T) &= p_0(x,y,t,T) + \epsilon^2 p_1(x,y,t,T) + \epsilon^4 p_2(x,y,t,T) + \hdots,\\
u(x,y,t,T) &= u_0(x,y,t,T) + \epsilon^2 u_1(x,y,t,T) + \epsilon^4 u_2(x,y,t,T) + \hdots,\\
v(x,y,t,T)  &= v_0(x,y,t,T) + \epsilon^2 v_1(x,y,t,T) + \epsilon^4 v_2(x,y,t,T) +\hdots.
\end{align}
\end{subequations}
 The initial condition of the solute concentration is $c(t=T=0)=\exp(-x^2)$.
First, we need to obtain the leading-order velocity and pressure solutions. These can be obtained by substituting equations (\ref{perturbation}) into equations (\ref{eq:non-2dim}), which gives
\begin{subequations}\label{leading_order}
\begin{gather}
    0 = \frac{\partial^2 u_0}{\partial y^2} + \frac{\partial p_0}{\partial x},\\
    \frac{\partial p_0}{\partial y} = 0,\\
    \frac{\partial u_0}{\partial x} + \frac{\partial v_0}{\partial y} = 0,\\
    \frac{\partial c_0}{\partial T}=\frac{\partial^2 c_0}{\partial y^2},
\end{gather}
\end{subequations}
to leading order. Considering both the initial condition and the no-flux conditions at the channel walls, i.e., $\frac{\partial c_0}{\partial y}(y=\pm 1/2)=0$, it must be true that $c_0(x,y,t,T)=c_0(x,t)$, with $c_0(t=0)=\exp(-x^2)$. Furthermore, equation (\ref{leading_order}$b$) indicates that $p_0$ is only a function of $x$ and $t$.

To obtain the leading-order velocities, we first take equation (\ref{leading_order}$a$) and solve for $u_0$, which is subject to the slip boundary condition (\ref{slip_condition}). This gives
\begin{equation}\label{u0_notfinished}
    u_0 = \frac{\Gamma_w}{D_s c_0}\frac{\partial c_0}{\partial x}+\frac{1}{8}(-1+4y^2)\frac{\partial p_0}{\partial x}.
\end{equation}
Here, $u_0$ has a term that includes $p_0(x,t)$, which can be found by considering the conservation of mass and integrating over the channel cross-section $A$, i.e.,
\begin{equation}\label{conser_mass}
    \frac{1}{A}\iint_A u_0 \,dA = 0.
\end{equation}
Following this approach, $p_0(x,t)$, is found to be
\begin{equation}\label{p0}
    p_0(x,t) = \frac{12\Gamma_w}{D_s}\ln c_0.
\end{equation}
With this result, $u_0$ can be simplified and written as
\begin{equation} \label{u_0}
    u_0(x,y,t) = \frac{\Gamma_w}{2D_s}\frac{\partial \ln c_0}{\partial x}(-1+12y^2).
\end{equation}
To solve for $v_0(x,y,t)$, we integrate the continuity equation (\ref{leading_order}$c$) and apply the boundary condition given by (\ref{bc:v}). This gives the leading-order $y$-component of velocity to be
\begin{equation}\label{v0}
    v_0(x,y,t) =  \frac{\Gamma_w}{2D_s}\frac{\partial^2 \ln c_0}{\partial x^2}(1-4y^2)y.
\end{equation}
As mentioned, the equivalent results for the flow in a cylindrical pipe can be found in Appendix \ref{Appendix:B}.

\subsection{Higher-order solute transport}\label{sec:c}
The higher-order solute concentration results from diffusioosmosis, which causes the deviation from pure diffusion. Using the leading-order velocity profiles found above, we seek a solution for the higher-order solute dynamics from (\ref{eq:non-2dim}$d$). 
Substituting our asymptotic expansion into the advection-diffusion equation (\ref{adv-diff}), we find, to leading-order, that
\begin{equation}\label{adv-diff-2nd}
    \frac{\partial c_1}{\partial T} + \frac{\partial c_0}{\partial t}+u_0\frac{\partial c_0}{\partial x} = \frac{\partial^2 c_0}{\partial x^2} + \frac{\partial^2 c_1}{\partial y^2}.
\end{equation}
The term involving $v$ has disappeared since $\frac{\partial c_0}{\partial y} = 0$. To find a solution to this problem, we first consider long times such that $T\gg 1$, but $t$ is finite. In this limit, $\frac{\partial c_1}{\partial T}$ is small. Then, averaging equation (\ref{adv-diff-2nd}) over the channel cross-section gives
\begin{equation}
    \frac{\partial c_0}{\partial t} = \frac{\partial^2 c_0}{\partial x^2}.
\end{equation}
The solution to this problem is given by
\begin{equation}\label{c0}
    c_0(x,t)=\frac{1}{\sqrt{1+4t}}\exp{\left(-\frac{x^2}{1+4t}\right)},
\end{equation}
which was previously developed by \citep{gupta2020diffusiophoresis}. The advection-diffusion equation (\ref{adv-diff-2nd}) can then be simplified to
\begin{equation}\label{adv-diff-3rd}
    \frac{\partial c_1}{\partial T} + \frac{\Gamma_w}{2D_s}\frac{\partial \ln c_0}{\partial x}(-1+12y^2)\frac{\partial c_0}{\partial x}=\frac{\partial^2 c_1}{\partial y^2},
\end{equation}
subject to the initial condition $c_1(t=T=0)=0$. To solve this, we note that at long times the fast-time dynamics should have all decayed such that the time derivative term can be ignored, and the equation can be integrated to yield
\begin{equation}\label{cinfty}
    c_1(T\rightarrow\infty)\sim c_1^\infty(x,y,t)=\frac{y^2}{4c_0}\frac{\Gamma_w}{Ds}(-1+2y^2)\left(\frac{\partial c_0}{\partial x}\right)^2+B(x,t)
\end{equation}
where $B(x,t)$ is a yet unknown function that results from the integration. To solve for $B(x,t)$, we substitute (\ref{cinfty}) into (\ref{adv-diff}), and take the cross-sectional average of the equation, which gives
\begin{equation}\label{pde_A}
    \frac{\partial^2 B}{\partial x^2} =\frac{\Gamma_w}{D_s}\frac{e^{-\frac{-x^2}{1+4t}}}{420(1+4t)^{9/2}}\left[49(1+4t)^2+48\frac{\Gamma_w}{D_s}(1+4t)x^2-32\frac{\Gamma_w}{D_s}x^4 \right]+\frac{\partial B}{\partial t}.
\end{equation}
Note that, until now, we have left the results in terms of a general $c_0$, but here and in the solution for $B$ below, we have substituted the specific solution for $c_0$ shown above since it is needed to solve the equation (\ref{pde_A}) using the Fourier transform approach. This procedure can be repeated for other arbitrary initial conditions as needed. Using a Fourier transform approach, $B(x,t)$ can be found to be
\begin{equation}
    B= \frac{\Gamma_w}{D_s}\frac{e^{-\frac{-x^2}{\alpha}}}{840\alpha^{9/2}}\left(49(\alpha x)^2+16\frac{\Gamma_w}{D_s}t\left(3\alpha^2-12\alpha x^2+4x^4 \right)-12\frac{\Gamma_w}{D_s}\alpha^2(\alpha-2x^2)\ln(\alpha)\right),
\end{equation}
where $\alpha(t)=1+4t$.  Note that $c_1^\infty$ only depends on the slow time $t$ and not the fast time $T$. It does not satisfy the initial condition, so it is not yet the full solution for the higher-order solute dynamics. We seek a solution of the form $c_1(x,y,t,T) = c_1^\infty(x,y,t)+\hat{c}_1(x,y,t,T)$. Substituting into equation (\ref{adv-diff-3rd}), we find
\begin{equation}
    \frac{\partial \hat{c}_1}{\partial T} = \frac{\partial^2 \hat{c}_1}{\partial y^2},~~~~~\text{with the initial condition}~~~~\hat{c}_1(T=0)=-c_1^\infty(t=0),
\end{equation}
the solution to which is
\begin{equation}\label{chat}
    \hat{c}_1 = \left. \frac{\Gamma_w}{4c_0D_s}\left(\frac{\partial c_0}{\partial x}\right)^2\right|_{t=0}\sum_{n=1}^\infty\frac{6(-1)^n}{n^4\pi^4}e^{-(2n\pi)^2T}\cos 2n\pi y.
\end{equation}
We can then construct a composite solution $c_1=c_1^\infty+\hat{c}_1$, which is valid for all $t$. The solution of $c_1$ can be verified to satisfy the conservation of mass by considering
\begin{equation}
    \int^\infty_{-\infty}\int^{1/2}_{-1/2} c_1(x,y,t,T)\mathrm{dy\,dx} = 0.
\end{equation}
Once again, analogous results for the coupled dynamics in a cylindrical pipe geometry can be found in Appendix \ref{Appendix:B}.

\subsection{Effective diffusivities}\label{sec:ave}
As mentioned, the diffusioosmotic slip flow at the channel walls induces a recirculating flow that drives an advective transport of the solute, altering the effective diffusivity of its transport as it diffuses along the channel. Here, we seek to characterize the effective diffusivity of this transport by deriving a 1D transport equation for the cross-sectionally averaged solute transport. This approach is analogous to that of \citet{taylor1953dispersion} and \citet{aris1956dispersion} in their famous work on solute dispersion in the presence of pressure-driven shear flow (see also, e.g., \citet{alessio2022diffusioosmosis,aminian2016boundaries}).

To begin, we define $c'(x,y,t)$ as the deviation of the solute concentration from its cross-sectionally averaged value $\overline{c(x,y,t)}$, where overbars are used to denote the average over a cross-section,
\begin{equation}\label{eq:pert}
    c'(x,y,t) = c(x,y,t) - \overline{c(x,t)}.
\end{equation}
Substituting this definition into the solute advection-diffusion equation gives
\begin{equation}\label{eq:pert_eq}
    \frac{\partial \overline{c}}{\partial t}+\frac{\partial c'}{\partial t}+u \frac{\partial c'}{\partial x} +u \frac{\partial \overline{c}}{\partial x}+ v \frac{\partial c'}{\partial y}=\frac{\partial^2 \overline{c}}{\partial x^2} + \frac{\partial^2 c'}{\partial x^2}+\frac{1}{\epsilon^2}\frac{\partial^2 c'}{\partial y^2}.
\end{equation}
Next, averaging this equation over the cross-section gives
\begin{equation}\label{eq:ave_eq}
    \frac{\partial \overline{c}}{\partial t} + \overline{u \frac{\partial c'}{\partial x}} + \overline{v \frac{\partial c'}{\partial y}} = \frac{\partial^2 \overline{c}}{\partial x^2}.
\end{equation}
Here, substituting in our solutions for $u$ and $v$ from above, this can be rewritten into a 1D diffusion equation with a known forcing term given by
\begin{equation}\label{1dforce}
    \frac{\partial \overline{c}}{\partial t} + \frac{\partial}{\partial x}\left[\left(\frac{\Gamma_w}{D_s}\right)^2\frac{\epsilon^2}{210}\left(\frac{\partial}{\partial x}\ln c_0 \right)^2\frac{\partial c_0}{\partial x} \right] = \frac{\partial^2 \overline{c}}{\partial x^2}.
\end{equation}
This equation can easily be solved numerically. For the purposes of making an analogy to the classic Taylor dispersion problem, this can be rewritten into a 1D pure diffusion problem by recognizing that $\frac{\partial c_0}{\partial x}-\frac{\partial \overline{c}}{\partial x}=\mathcal{O}(\epsilon^2)$, which gives
\begin{equation}\label{eq:1D_dis}
    \frac{\partial \overline{c}}{\partial t} = \frac{\partial}{\partial x}\left(D_\textrm{eff} \frac{\partial \overline{c}}{\partial x}\right),
\end{equation}
where the effective diffusivity $D_\textrm{eff}$ is given by
\begin{equation}\label{eq:deff}
    D_\textrm{eff} = 1+\left(\frac{\Gamma_w}{D_s}\right)^2\frac{\epsilon^2}{210}\left(\frac{\partial}{\partial x}\ln c_0\right)^2 +\mathcal{O}(\epsilon^4).
\end{equation}

From equation (\ref{eq:deff}) we see that to leading-order the effective nondimensional diffusivity is $D_\text{eff}=1$, and the effects of diffusioosmosis on the dispersion are $\mathcal{O}(\epsilon^2)$. That is, as $\epsilon\rightarrow0$, the initial condition of the solute plug becomes more spread out, the concentration gradients get weaker, and the diffusioosmosis becomes negligible. The same behavior occurs as $t\rightarrow\infty$ as the solute spreads out over long times. Curiously, the contribution from diffusioosmosis is also $\mathcal{O}((\Gamma_w/D_s)^2)$. Thus, despite the non-linearity of the diffusioosmotic boundary condition, flipping the sign of $\Gamma_w$ (and thus the direction of the recirculation) results in the same effective diffusivity. This will be made slightly more surprising when we visualize the results below and see that the deviations in the solute concentration from $c_0$ are not mirror images of each other when the sign of the mobility is flipped.

Following a similar approach and using the results from Appendix \ref{Appendix:B}, the analogous averaged transport equation for the cylindrical pipe case is given by
\begin{equation}\label{1dforce_cyl}
    \frac{\partial \overline{c}}{\partial t} + \frac{\partial}{\partial z}\left[\left(\frac{\Gamma_w}{D_s}\right)^2\frac{\epsilon^2}{48}\left(\frac{\partial}{\partial z}\ln c_0 \right)^2\frac{\partial c_0}{\partial z} \right] = \frac{\partial^2 \overline{c}}{\partial z^2}.
\end{equation}
This can also be written into a form analogous to Taylor dispersion as
\begin{equation}
    \frac{\partial \overline{c}}{\partial t} = \frac{\partial}{\partial z}\left(D_\textrm{eff} \frac{\partial \overline{c}}{\partial z}\right),
\end{equation}
where the effective diffusivity $D_\textrm{eff}$ to leading-order is given by
\begin{equation}
    D_\textrm{eff} = 1+\left(\frac{\Gamma_w}{D_s}\right)^2\frac{\epsilon^2}{48}\left(\frac{\partial}{\partial z}\ln c_0\right)^2+\mathcal{O}(\epsilon^4).
\end{equation}
Note that the contribution of diffusioosmosis to the effective diffusivity in a 2D channel flow is a factor of 48/210 weaker than in a cylindrical pipe system. This is a consequence of the fact that for a given slip velocity at the walls, the centerline velocity in a cylindrical pipe must be greater than in a 2D channel flow, leading to greater velocity gradients, greater distortion of the solute profile, and a greater contribution to the effective diffusivity enhancement. Using these effective diffusivities along with the 1D diffusion equation, the evolution of the cross-sectionally averaged solute concentration is quite efficient to compute numerically.

\section{Numerical methods}\label{sec:numerical}
To validate the theoretical results above, we performed numerical simulations for the coupled transport in both geometries. In this section, we describe the numerical methods used and show that the theoretical results above accurately match the numerical results. The velocity profiles are solved for the 2D channel flow case using a Fourier transform approach and for the cylindrical pipe flow case using a multigrid relaxation approach. Because we are dealing with the low Reynolds number regime and the fluid dynamics can be treated as quasi-steady, the numerical approach can be outlined as follows. First, we initialize the solute concentration to the previously-described Gaussian distribution. We then solve for the quasi-steady velocity profile using either a numerical multigrid relaxation approach or the theoretical Fourier transform approach. Next, using the determined velocity profiles, we update the solute concentration profile using a numerical solution of the governing advection-diffusion equation. This procedure is then repeated until the final time.

\subsection{Velocity solver}\label{sec:sim_vel}
For the case of incompressible Newtonian Stokes flow, the governing equations can be simplified to a biharmonic equation for the streamfunction, $\psi$, which is given by
\begin{equation}
    \nabla^4 \psi = 0.
\end{equation}
This is a convenient formulation, as it allows for the solution of the velocity profiles without needing to solve for the pressure. In the Cartesian coordinate system, a Fourier transform approach can be used to greatly accelerate the numerical solution of this equation. In particular, the biharmonic equation can be Fourier transformed in the $x$ direction as
\begin{equation}\label{FT}
    k^4\hat{\psi}(k,y)-2k^2\frac{\partial^2  \hat{\psi}(k,y)}{\partial y^2}+\frac{\partial^4 \hat{\psi}(k,y)}{\partial y^4}=0.
\end{equation}
Here, we ignore the functional dependence of $\psi$ on time because the velocity can be treated as quasi-steady. The boundary conditions for (\ref{FT}) are $\hat{\psi}(k,0)=0$, $\frac{\partial^2 \hat{\psi}}{\partial y^2}\bigr|_{y = 0} = 0$, $\hat{\psi}(k,\pm 1/2)=0$, and $\frac{\partial \hat{\psi}}{\partial y}\bigr|_{y=\pm 1/2}=\hat{u}_\textrm{wall}$. Here, $\hat{u}_\textrm{wall}$ can be found by using the Fast Fourier Transform \texttt{fft()} in MATLAB, and $\psi$ can be found by using the inverse Fast Fourier Transform of $\hat{\psi}$ using \texttt{ifft()}. Finally, the velocity components can be obtained directly from $\psi$ using
\begin{equation}
    u = \frac{\partial \psi}{\partial y},~~~~~~v = -\frac{\partial \psi}{\partial x}.
\end{equation}

For the case in cylindrical coordinates, we have not found a similar approach to use a Fourier transform to rapidly solve the biharmonic equation, so we instead do this directly using a multigrid solution to a set of coupled Poisson equations, and we then use the direct finite difference method to solve for the velocity. In particular, the biharmonic equation can be written as follows
\begin{subequations}\label{poisson_cyl}
\begin{align}
    D^2 \psi &= r\frac{\partial}{\partial r}\left( \frac{1}{r}\frac{\partial \psi}{\partial r}\right) + \frac{\partial^2 \psi}{\partial z^2}=\phi,\\
    D^2 \phi &= r\frac{\partial}{\partial r}\left( \frac{1}{r}\frac{\partial \phi}{\partial r}\right) + \frac{\partial^2 \phi}{\partial z^2}=0,
\end{align}
\end{subequations}
where $D^2$ is a special operator in cylindrical coordinates that is useful for the biharmonic equation, which is given by $D^2 = r\frac{\partial}{\partial r}\left( \frac{1}{r}\frac{\partial}{\partial r}\right) + \frac{\partial^2}{\partial z^2}$ \citep{stimson1926motion,brenner1961slow}. The wall has boundary conditions of $\frac{1}{r}\frac{\partial \psi}{\partial r} = u_\text{wall}$ and $\phi = -\frac{1}{r}\frac{\partial\psi}{\partial r} + \frac{\partial^2 \psi}{\partial r^2}$. All of the other boundaries have $\psi = 0$ and $\phi = 0$ because of symmetry conditions and no-flux conditions at the end of the channel.

This set of coupled Poisson equations is solved by using the Gauss-Seidel relaxation method with second-order accuracy in space and time for both governing equations and boundary conditions. As mentioned, the multigrid approach was used to rapidly accelerate the convergence of this iterative approach.

\subsection{Concentration solver}
The solute concentration profiles can be numerically solved using the advection-diffusion equation. We use finite difference with the approximate factorization method to solve the advection-diffusion equation for both geometries \citep{moin2010fundamentals}. The semi-discretized governing equations applicable to the approximate factorization method for the Cartesian and cylindrical coordinate systems are given, respectively, by
\begin{equation} \label{adv_ap_cart}
    \begin{split}
        \left[I-\frac{D_s \Delta t}{2}\delta_{xx} \right]\left[I -\frac{D_s \Delta t}{2}\delta_{yy} \right]c^{n+1} = -\frac{\Delta t}{2}\left(3u^n\frac{\partial c^n}{\partial x} + u^{n-1}\frac{\partial c^{n-1}}{\partial x}\right) \\
        -\frac{\Delta t}{2}\left(3v^n\frac{\partial c^n}{\partial y} + v^{n-1}\frac{\partial c^{n-1}}{\partial y}\right) + \left[I+\frac{D_s \Delta t}{2}\delta_{xx} \right]\left[I +\frac{D_s \Delta t}{2}\delta_{yy} \right]c^{n},
    \end{split}
\end{equation}
and
\begin{equation}\label{adv_ap_cyl}
\begin{split}
    \left[I-\frac{D_s\Delta t}{2}\left(\delta_{rr}+\frac{1}{r}\delta_r\right) \right]\left[I -\frac{D_s\Delta t}{2} \delta_{zz}\right]c^{n+1} = -\frac{\Delta t}{2}\left(3u_r^n\frac{\partial c^n}{\partial r}-u_r^{n-1}\frac{\partial c^{n-1}}{\partial r}\right) \\
    -\frac{\Delta t}{2}\left(3u_z^n\frac{\partial c^n}{\partial z}-u_z^{n-1}\frac{\partial c^{n-1}}{\partial z}\right) + \left[I+\frac{D_s\Delta t}{2}\left(\delta_{rr}+\frac{1}{r}\delta_r\right) \right]\left[I +\frac{D_s\Delta t}{2} \delta_{zz}\right]c^{n},
\end{split}
\end{equation}
where $\Delta t$ is the timestep, $n$ indicates the current time index, $I$ represents the identity matrix, and the $\delta_{xx}$, etc. are the spatial derivative matrices. Here, implicit second-order accurate time-stepping is used for the diffusive terms, and second-order Adams-Bashforth is used for the advective terms. For the cylindrical case, second-order finite differencing was used for the spatial derivatives, whereas fourth-order finite differencing was uses for the Cartesian case. To start the simulation, we first take a series of very small timesteps using a first-order Euler method to calculate the advective terms. For each time step, the solute concentration profiles are updated in time by solving equations (\ref{adv_ap_cart}) or (\ref{adv_ap_cyl}), and then the fluid velocities are recalculated as described above for use in the next timestep. Note that in all cases we add a small offset background concentration of $10^{-7}$ to prevent the so-called `ballistic motion' described by \cite{gupta2020diffusiophoresis} which represents the background ion concentration typically present in solution due to dissolved CO$_2$ or other factors. In this section, we have developed and presented the numerical methods used for both systems. In the following section, we explore the range of results and the physical evolution of such systems with numerical validation, and we show that the cross-sectionally averaged approach closely approximates the results of fully 2D simulations and can greatly simplify the analysis as in the case of Taylor dispersion.

\section{Results}\label{sec:result}
\begin{figure}
    \centering
    \includegraphics[width=\textwidth]{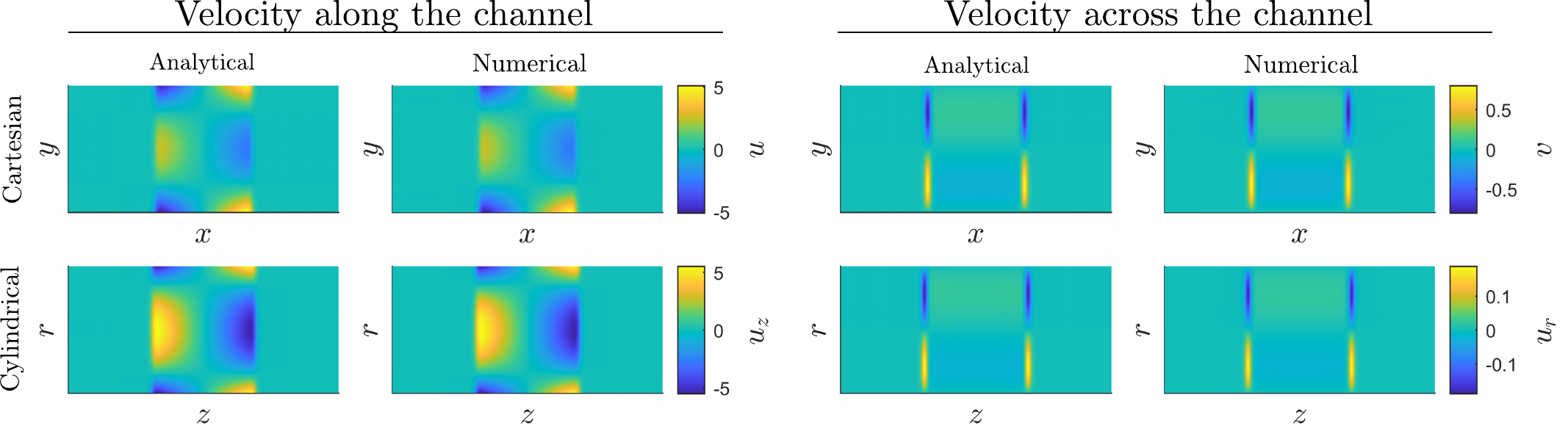}
    \caption{The comparison between numerical and theoretical velocity predictions in both coordinate systems. The recirculating velocity is due to the diffusioosmotic motion at the channel walls, which is driven by the $u_{\text{wall}} = \frac{\Gamma_w}{D_s}\frac{\partial \ln c}{\partial x}$ or $u_{\text{wall}} = \frac{\Gamma_w}{D_s}\frac{\partial \ln c}{\partial z}$ slip boundary condition for the Cartesian and cylindrical coordinate systems, respectively. Results are computed for $\Gamma_w/D_s = -1$ and $\epsilon=0.1$ at $t=0.2$.}
    \label{fig:cyl_vel}
\end{figure}

To begin, we first use the numerical simulations to validate the theoretical predictions. An example comparison between the numerical and theoretical results is shown in figure \ref{fig:cyl_vel}.
The analytical predictions of velocities in the parallel-plate channel are calculated by using equations (\ref{u_0}) and (\ref{v0}), and those in the cylindrical channel are calculated from (\ref{eq:uz0}) and (\ref{eq:ur0}). Results are compared for $\Gamma_w/D_s = -1$ and $\epsilon=0.1$ at $t=0.2$. With a negative diffusioosmotic mobility, the wall slip velocity is away from the peak solute concentration, driving flow away from the centerplane ($x=0$ or $z=0$) at the walls and toward the centerplane along the channel centerline ($y=0$ or $r=0$). The theoretical and numerical results agree well. One feature of the results to notice is that the velocity along the centerline in the cylindrical case is enhanced relative to the Cartesian case. We will see later how this alters the effective disperion in such configurations.
\begin{figure}
    \centering
    \includegraphics[width=\textwidth]{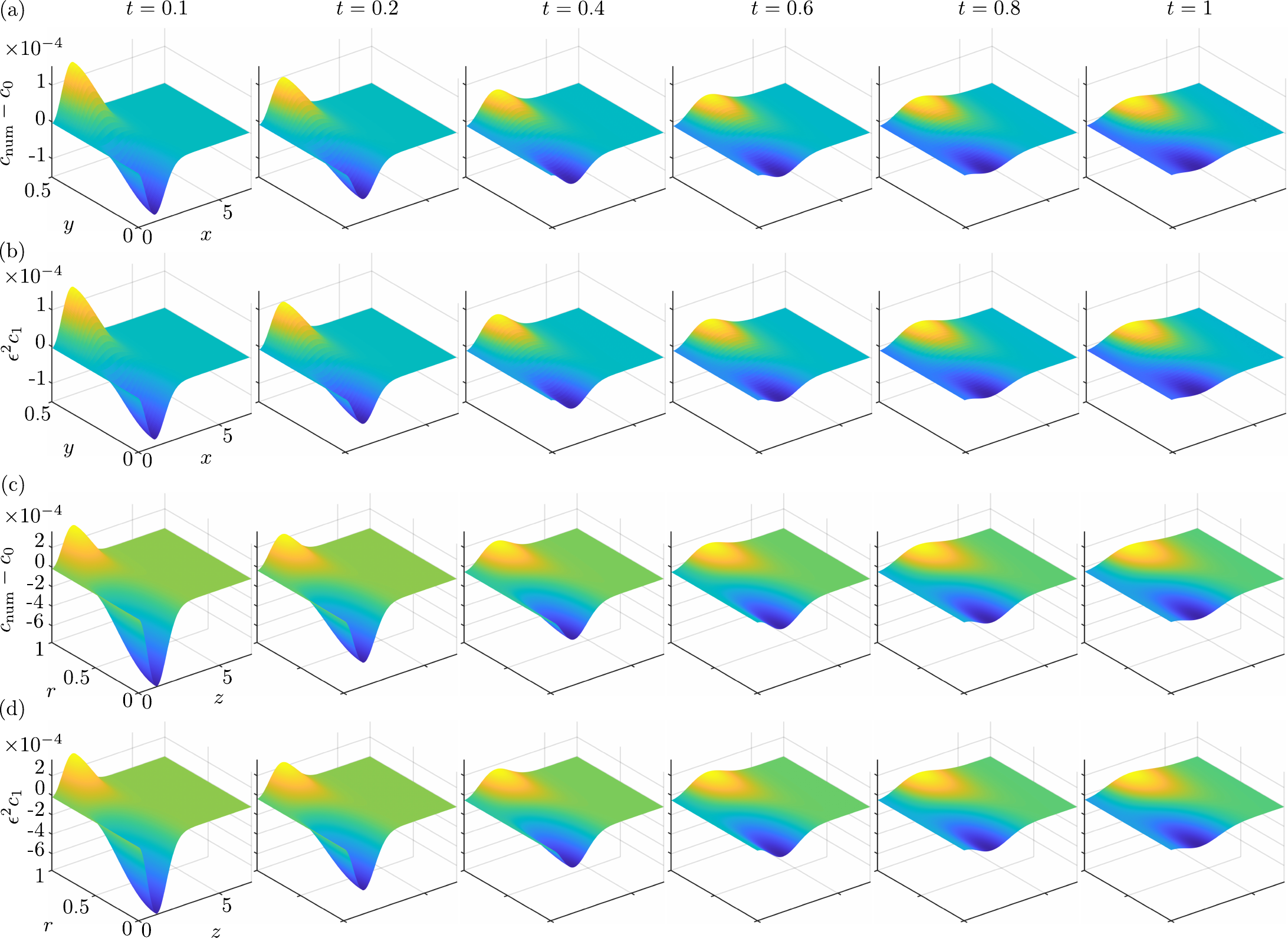}
    \caption{Evolution of the higher-order solute concentration in both the Cartesian (a,b) and cylindrical (c,d) geometries. This illustrates the deviation of the solute concentration profile from the purely 1D dynamics and represents the role of the diffusioosmotic dispersion. The panels with $c_\text{num}-c_0$ represent the numerically computed solute evolution minus the theoretical 1D solution, and panels with $\epsilon^2 c_1$ show the theoretically calculated higher-order solute profile. Results are presented over time for $\Gamma_w/D_s = -1$ and $\epsilon=0.1$.}
    \label{fig:c1_comp}
\end{figure}
Figure \ref{fig:c1_comp} shows the comparison between (a,c) numerical simulations and (b,d) theoretical predictions of the higher-order solute concentration in the parallel-plate (a,b) and the cylindrical (c,d) channels, respectively. Here, the results are plotted over one quarter of the domain due to symmetry. The comparison is again calculated with $\Gamma_w/D_s = -1$ and $\epsilon = 0.1$, and the results demonstrate that the numerical results closely match the theoretical predictions.

\begin{figure}
\centering
\includegraphics[width=0.8\textwidth]{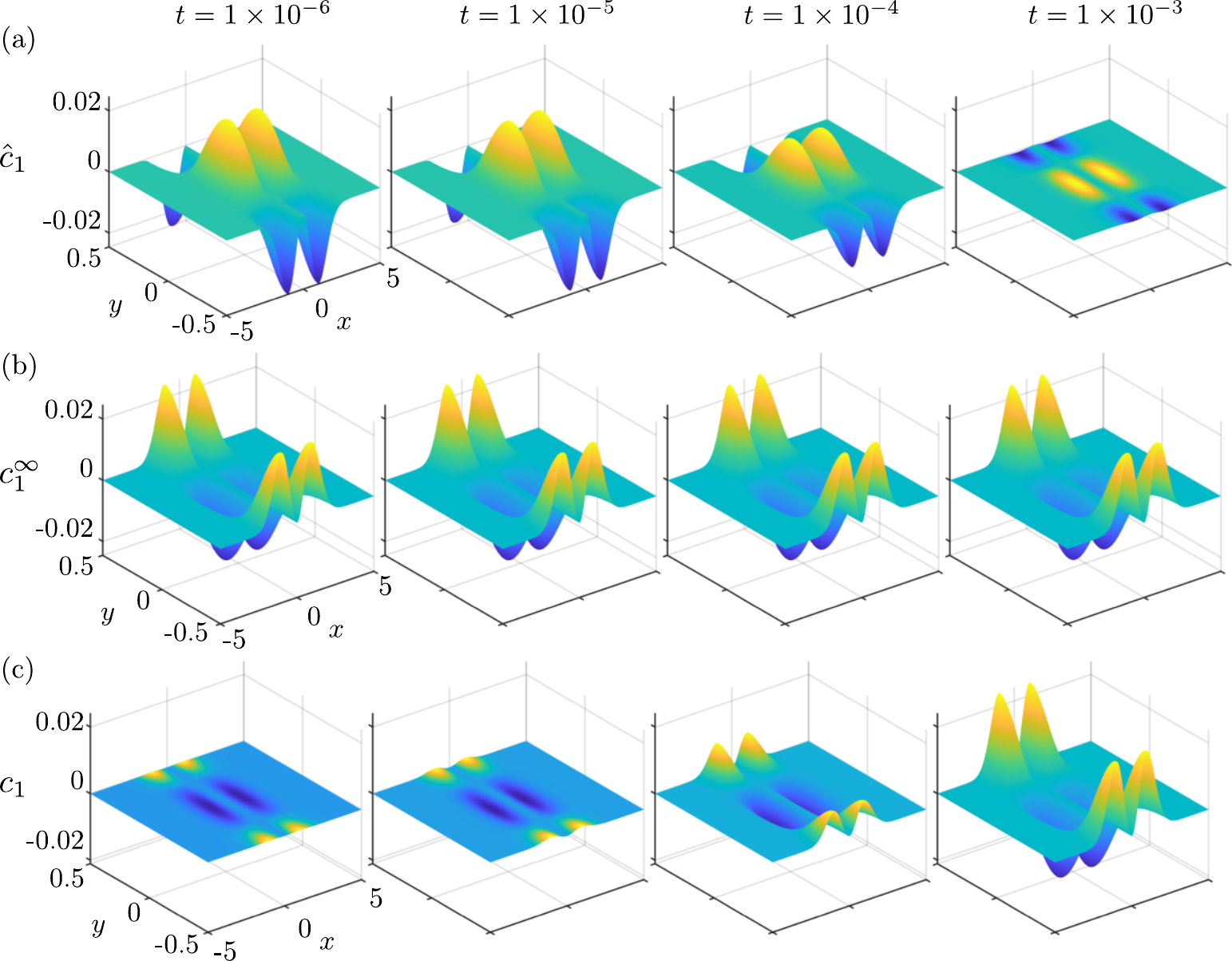}
\caption{Components of the higher-order solute contribution during the early-time regime. Results are calculated for $\Gamma_w/D_s = -1$ and $\epsilon = 0.1$ up to a time of $t=1\times10^{-3}$. Here, $c_1^\infty$ is calculated from equation (\ref{cinfty}) and corresponds to the long-time solution from the multiple timescale analysis. The $\hat{c}_1$ component is calculated from equation (\ref{chat}) and corresponds to the fast-time dynamics that are required to satisfy the initial condition. The total higher-order solute profile is then given by $c_1 = c_1^\infty+\hat{c}_1$. The contribution due to the fast-time dynamics decays over the timescale for solute diffusion across the channel, and the long-time contribution decays over the timescale for diffusion along the channel.}
    \label{fig:fasttime}
\end{figure}

Having validated the theoretical solutions using numerical simulations, we now provide a detailed examination of the diffusioosmotic dispersion process. We first consider the early-time dynamics of the system over which the initial deviation of the solute profile from the purely 1D dynamics forms. Figure~\ref{fig:fasttime} illustrates these early time dynamics by presenting visualizations of both $\hat{c}_1$, $c_1^\infty$, and $c_1$ in the early-time regime for times up to $t=1\times10^{-3}$ with $\epsilon=0.1$ and $\Gamma_w/D_s = -1$. 
Recall that the fast timescale corresponds to the characteristic time for diffusion to occur across the channel and is represented by $\hat{c}_1$ from equation (\ref{chat}). In contrast, the slow timescale corresponds to the diffusion along the channel and is represented by $c_1^\infty$ from equation (\ref{cinfty}). Here, $c_1$ is the total deviation of the solute concentration profile from the purely 1D dynamics and is formed by the sum of both $c_1^\infty$ and $\hat{c_1}$. As can be seen, the purpose of $\hat{c}_1$ is to cancel out the initial condition of $c_1^\infty$ such that the initial condition of $c_1$ can be zero. Then, in this example, $\hat{c}_1$ has almost entirely decayed by $t=10^{-3}$ after which the solution is dominated by the slow-time dynamics.

\begin{figure}
\centering
\includegraphics[width=\textwidth]{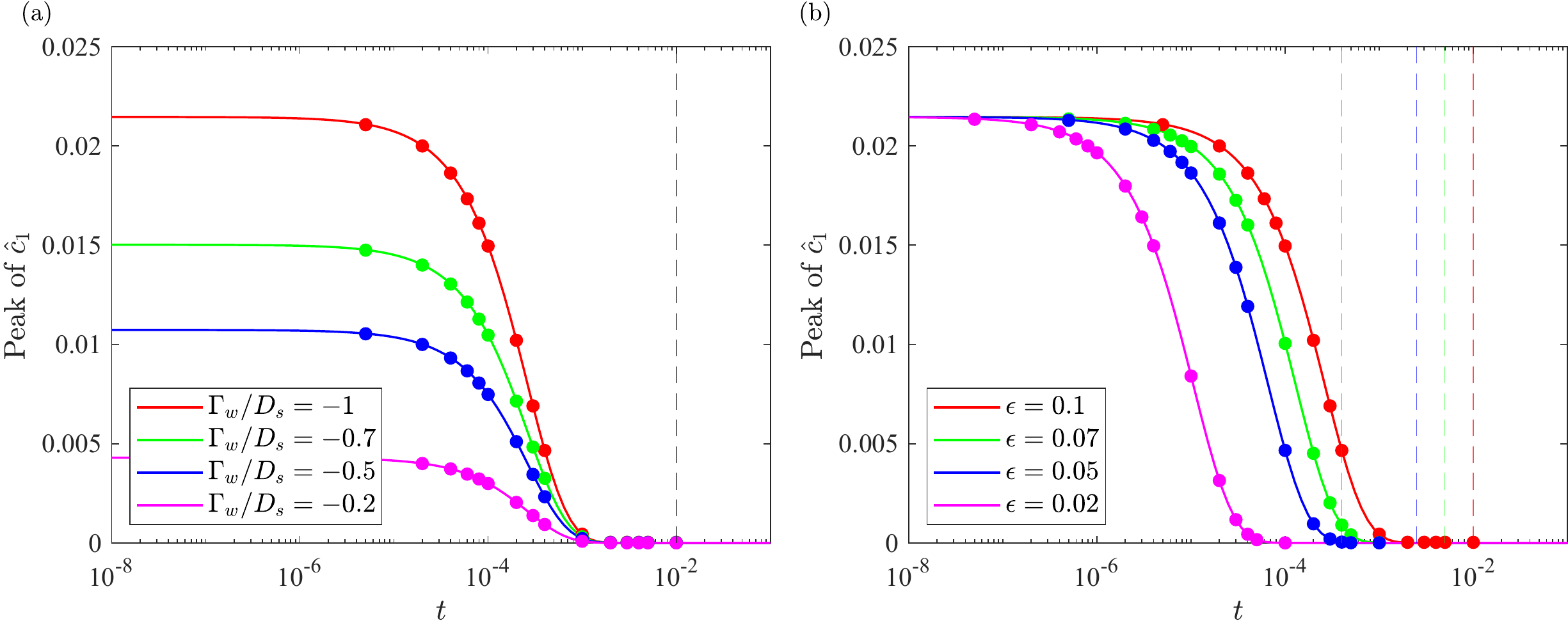}
\caption{Evolution of the peak values of $\hat{c}_1$ in the channel over time as functions of (a) $\Gamma_w/D_s$ for fixed $\epsilon=0.1$ and (b) $\epsilon$ for fixed $\Gamma_w/D_s=-1$. Solid dots indicate the 2D numerical simulation results, $\hat{c}_\textrm{num} = (c-c_0)/\epsilon^2-c_{1,\textrm{theory}}^\infty$. The theoretical predictions show an excellent agreement with the 2D numerical simulation. The dashed lines correspond to the time when $t=\epsilon^2$. Recall that $\hat{c}_1$ represents the fast-time dynamics in the system corresponding to solute diffusion across the channel and is expected to decay over the timescale $t\sim\epsilon^2$ as shown. In (a) the increased magnitude with increasing $\Gamma_w/D_s$ reflects the enhanced dispersion with stronger diffusioosmosis.}\label{fig:peak of c1}
\end{figure}

The early-time dynamics are expected to decay over the timescale $\epsilon^2$. This can be verified by plotting the peak value of $\hat{c}_1$ over a range of mobilities and $\epsilon$ values, which is shown in figure \ref{fig:peak of c1}. Solid dots correspond to the results of 2D simulations which are calculated as $\hat{c}_\textrm{num} = (c-c_0)/\epsilon^2-c_{1,\textrm{theory}}^\infty$.
Specifically, figure \ref{fig:peak of c1}$(a)$ shows the peak of $\hat{c_1}$ with $\epsilon = 0.1$ at various $\Gamma_w/D_s$ values. As can be seen, higher $\Gamma_w/D_s$ values correspond to larger peak $\hat{c}_1$ values, reflecting the enhanced dispersion in those cases. For all $\Gamma_w/D_s$, the peak $\hat{c}_1$ values have all apparently decayed before $t$ reaches $\epsilon^2=10^{-2}$. Figure \ref{fig:peak of c1}(b) extends these results by considering cases with different $\epsilon$ values at a constant $\Gamma_w/D_s=-1$. Here, the dashed lines are the locations where $t=\epsilon^2$. As expected, in every case the peak $\hat{c}_1$ value vanishes over the timescale $t=O(\epsilon^2)$ as predicted by the theory.

\begin{figure}
\centering
\includegraphics[width=\textwidth]{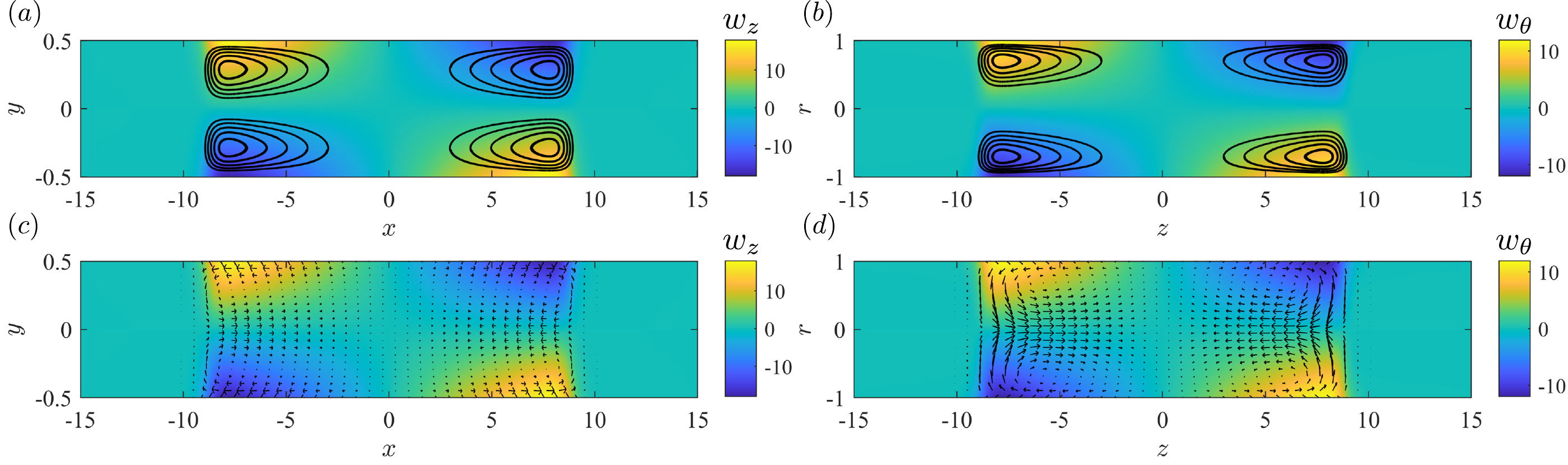}
\caption{Nondimensional vorticity and flow visualizations of the recirculation driven by diffusioosmosis for $\Gamma_w/D_s=-1$ at $t=1$. (a) and (c) correspond to the 2D channel flow case, and (b) and (d) correspond to the axisymmetric pipe flow case. Streamlines highlighting the recirculation zones are shown in (a) and (b), and velocity vector maps are shown in (c) and (d). Results correspond to the leading-order velocity profiles and thus are independent of $\epsilon$.}\label{fig:vorticity}
\end{figure}

In \S\ref{sec:theory}, we developed a theoretical model for the higher-order correction to the solute concentration profile due to diffusioosmosis. As discussed, the diffusioosmotic slip flow drives a recirculating flow in the channel that alters the transport of the solute. The vorticity and recirculating fluid flow in the channels are shown in figure \ref{fig:vorticity} at $t=1$ for both coordinate systems.
Here, (a) and (c) show the nondimensional vorticity on the cross-section for the parallel-plate and cylindrical channels, respectively, with $\Gamma_w/D_s=-1$. Streamlines illustrating the recirculation on the cross-section are superposed on top of the color map. Panels (b) and (d) show the same nondimensional vorticity data but with scaled velocity vector maps superposed instead. With, $\Gamma_w/D_s = -1$, the diffusioosmosis at the channel walls drives a slip flow away from the centerplane, driving a recirculating flow that is towards the centerplane along the channel centerline. The flow directions and the signs of the vorticity will be reveresed in cases with positive mobilities.

\begin{figure}
    \centering
    \includegraphics[width=\textwidth]{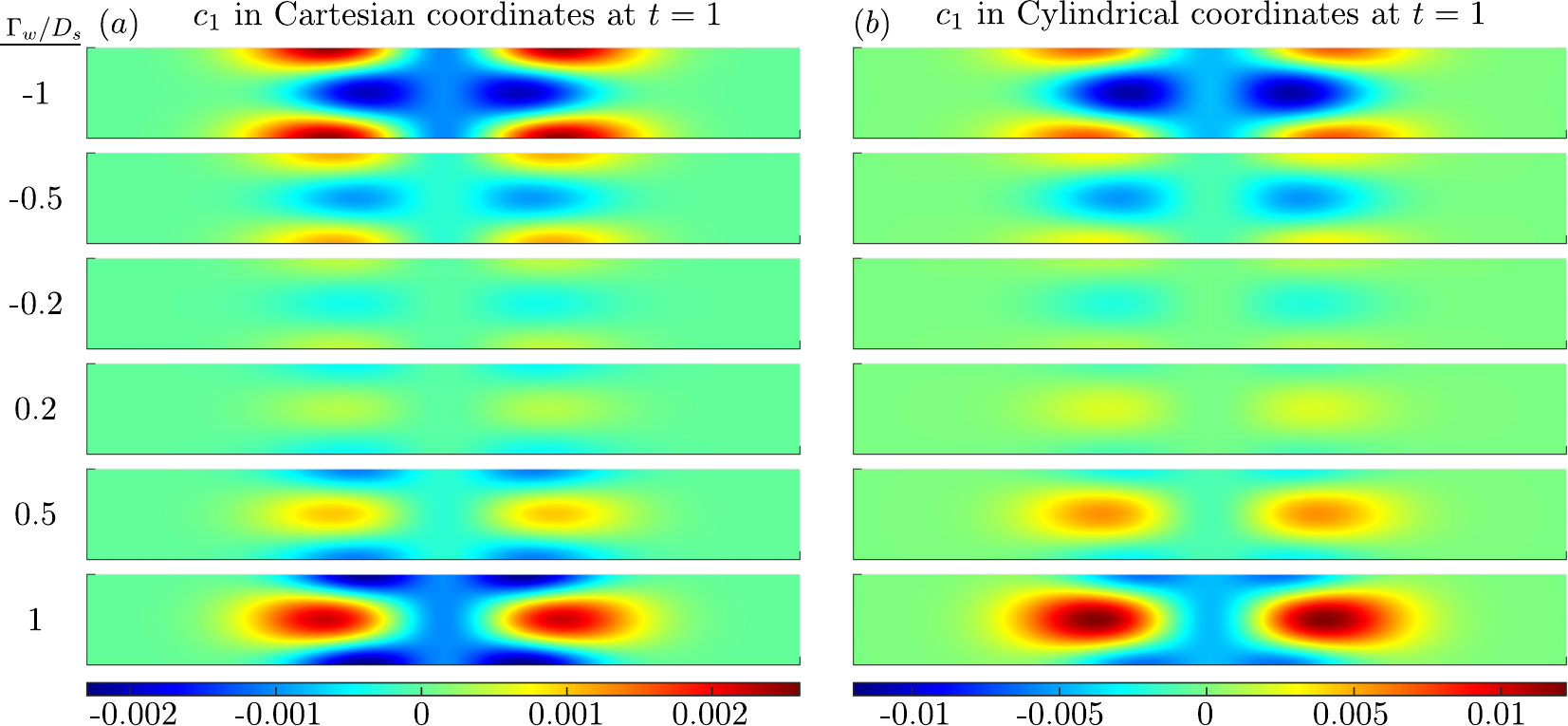}
    \caption{Higher-order solute concentration profiles $c_1$ as functions of $\Gamma_w/D_s$ at $t=1$. Panel (a) corresponds to the 2D Cartesian channel flow system and is calculated from equation (\ref{cinfty}), while panel (b) corresponds to the axisymmetric pipe flow case and is calculated from equation (\ref{c1infty_cyl}). In both panels, the vertical coordinate ($y$ or $r$) has been stretched by a factor of 2 for visualization purposes.}
    \label{fig:GwDs}
\end{figure}

Next, we visualize the higher-order solute dynamics on the cross-section to better understand the role of the diffusioosmotic dispersion on the solute transport. Figure \ref{fig:GwDs} shows the theoretical values of $c_1$ for both parallel-plate (a) and cylindrical (b) channels as functions of $\Gamma_w/D_s$ at a fixed time of $t=1$.
In order to interpret the figure, recall that negative $\Gamma_w$ values correspond to slip flow away from the centerplane along the walls and toward the centerplane along the centerline of the channel, while positive $\Gamma_w$ values correspond to flows that recirculate in the opposite direction. Regions of positive $c_1$ (red) indicate locations that have increased solute concentration relative to the 1D pure diffusion case ($c_0$), and regions of negative $c_1$ (blue) have relatively less concentration relative to $c_0$. We can interpret the formation of these regions as follows. For example, consider the case with $\Gamma_w/D_s=-1$ in the parallel-plate channel. Along the channel walls, the flow is away from the centerplane $x=0$, pulling the relatively higher concentration fluid away from $x=0$ and enhancing the solute concentration somewhat away from the centerplane. This is enhanced by the recirculating nature of the flow that also pulls flow away from the centerline of the channel and towards the walls. Along the centerline, the flow is towards the centerplane, pulling relatively lower concentration fluid toward the centerplane and resulting in a depletion region. These effects are flipped in the case of positive $\Gamma_w/D_s$. One last point to note is that the cylindrical case has relatively greater solute depletion and enhancement along the channel centerline due to the relatively greater centerline velocity in a cylindrical pipe compared to a 2D channel flow for the same wall slip velocity.

\begin{figure}
    \centering
    \includegraphics[width=\textwidth]{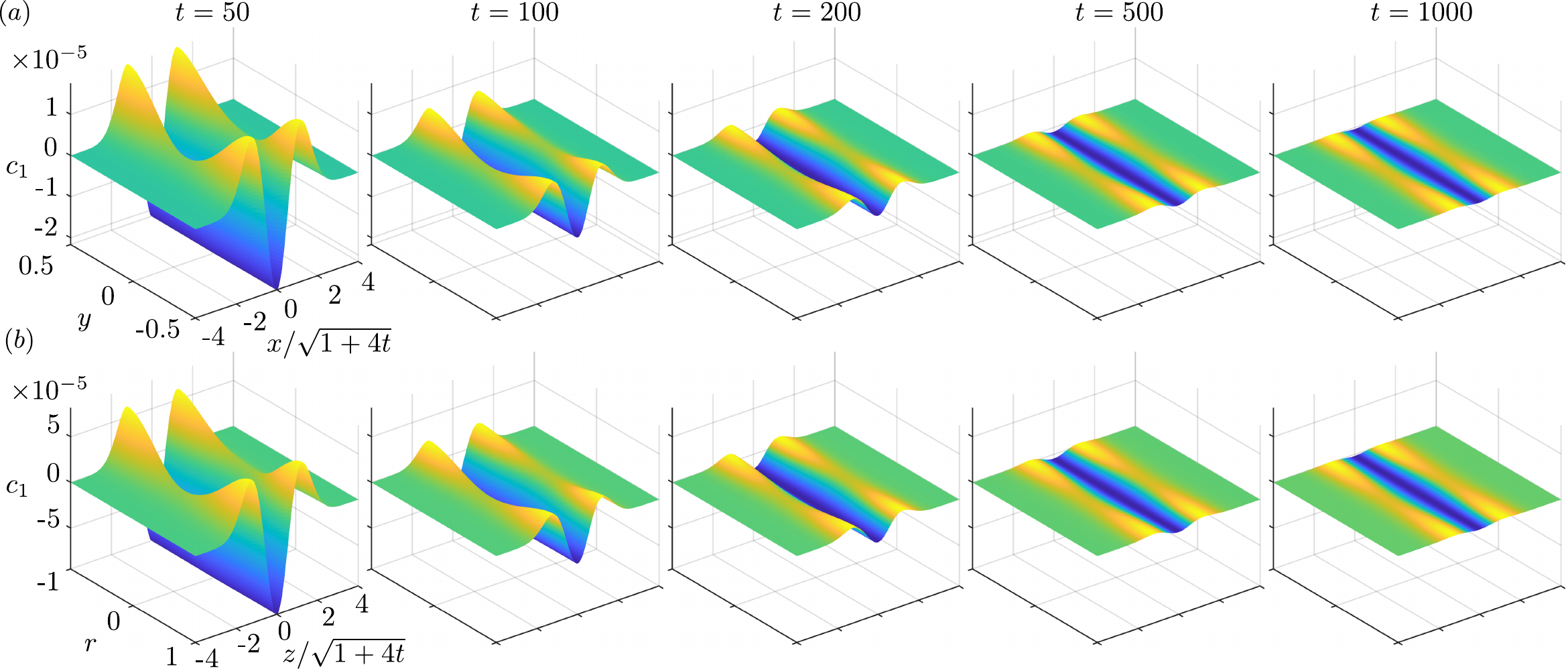}
    \caption{Long-time behavior of the higher-order solute concentration $c_1$ with $\Gamma_w/D_s=-1$ for times up to $t=1000$. Panel (a) corresponds to the 2D channel flow case, and panel (b) corresponds to the axisymmetric pipe flow case. In both cases, the axial coordinate is scaled by $\sqrt{1+4t}$, demonstrating that the higher-order solute effects spread at the same rate as $c_0$. As time proceeds, the solute concentration gradient at the walls decreases as the solute pulse spreads out, leading to decreased diffusioosmosis at the channel walls, less recirculation, and thus less dispersion. Ultimately, the higher-order profile smears out by diffusion, and the dynamics approach those of pure diffusion.}
    \label{fig:longtime}
\end{figure}

Next, we visualize the long-time behavior of the solute concentration profile as modeled by $c_1^\infty$. Recall that for $t>\mathcal{O}(\epsilon^2)$, the higher-order solute profile is $c_1\approx c_1^\infty$, as $\hat{c}_1$ has already decayed. The long-time results are shown in figure \ref{fig:longtime} for times up to $t=1000$ with $\Gamma_w/D_s=-1$. Here, the horizontal axis is scaled by $\sqrt{1+4t}$, since this represents the rate of spread of $c_0$ along the channel.
Recall that with our nondimensionalization, this corresponds to 1000 times the characteristic time for diffusion along the channel, such that the initial pulse of solute has well decayed by this time. As shown in the figure, over long times the higher-order solute profile also spreads significantly in the axial direction, retaining qualitatively the same shape, and ultimately decaying. As the initial pulse of solute decays, the concentration gradient at the walls likewise decays such that the diffusioosmosis and recirculating flow also decay with time leading to the ultimate decay of $c_1$. Here, panel (a) corresponds to the 2D channel flow case, and panel (b) corresponds to the axisymmetric pipe flow case. The only significant notable difference between the two cases is that the magnitude of the higher-order solute concentration is a factor of 4-5 higher for the cylindrical case. As mentioned, this is due to the relatively greater centerline velocity in the axisymmetric geometry, which yields greater velocity gradients and enhanced dispersion.

\begin{figure}
    \centering
    \includegraphics[width=\textwidth]{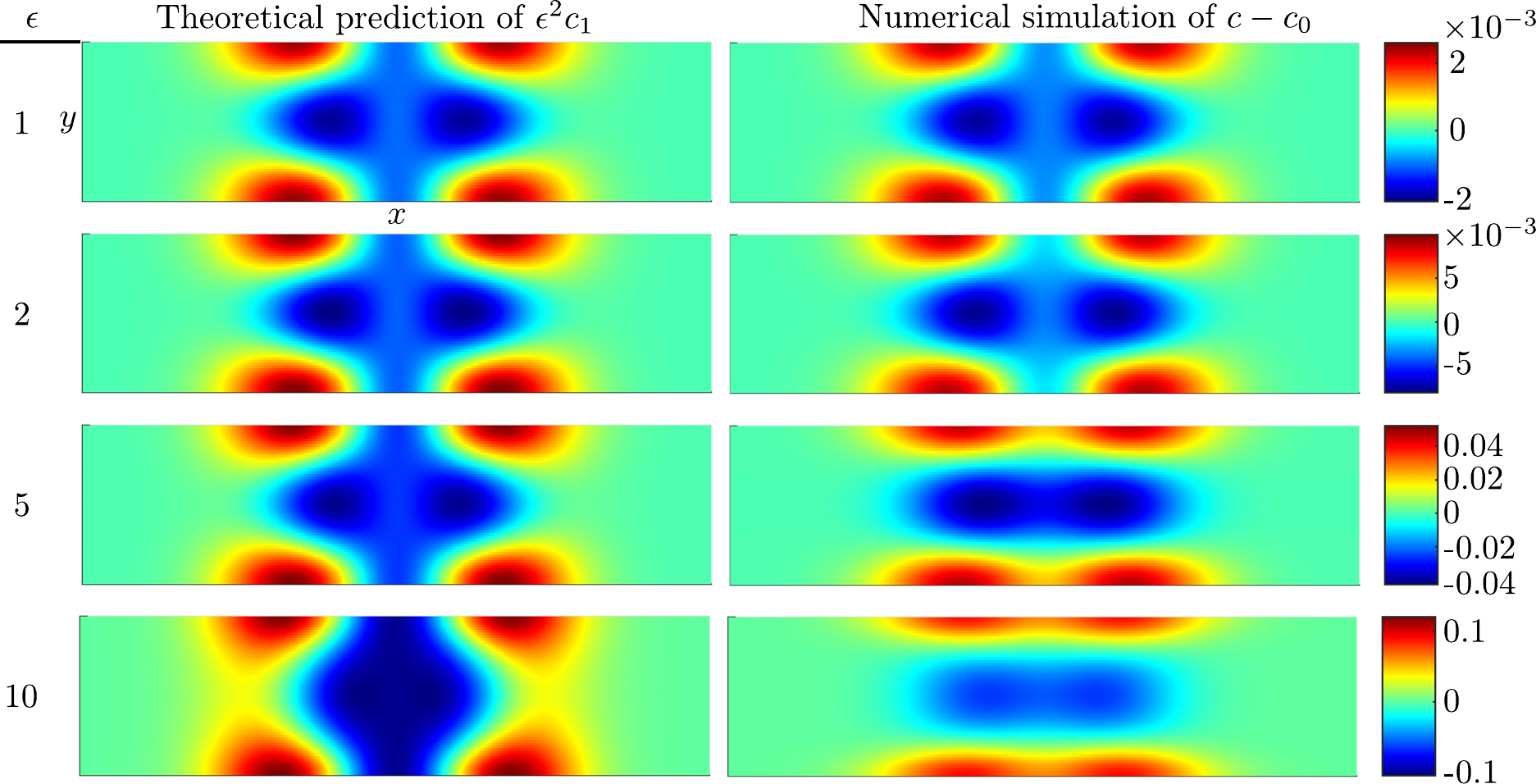}
    \caption{Higher-order solute concentration profiles $\epsilon^2c_1$ in the parallel-plate channel with $\Gamma_w/D_s=-1$ at $t=1$. Panel (a) corresponds to the theoretical predictions and is calculated from equation (\ref{cinfty}) and panel (b) corresponds to the numerical simulation results. As can be seen, the theoretical results appear to break down above approximately $\epsilon=2$.}
    \label{fig:varying_epsilon}
\end{figure}

Before proceeding to investigate the cross-sectionally averaged dynamics, we consider the effect of varying $\epsilon$ and the breakdown of the theoretical solution at large $\epsilon$. Figure \ref{fig:varying_epsilon} shows the theoretical and numerical predictions of the higher-order solute concentration profiles in the parallel-plate channel with $\Gamma_w/D_s=-1$ at $t=1$.
As shown above, the theoretical predictions show good agreement with the numerical results in the limit of small $\epsilon$. Here, we see that reasonable quantitative agreement between the theory and simulations exists up to about $\epsilon=2$, above which the theoretical predictions break down. In particular, as can be seen for $\epsilon=10$ in Figure \ref{fig:varying_epsilon}, the numerical results show a much more uniform depletion along the channel centerline near $x=0$ compared to the theoretical results.

\begin{figure}
\centering
\includegraphics[width=\textwidth]{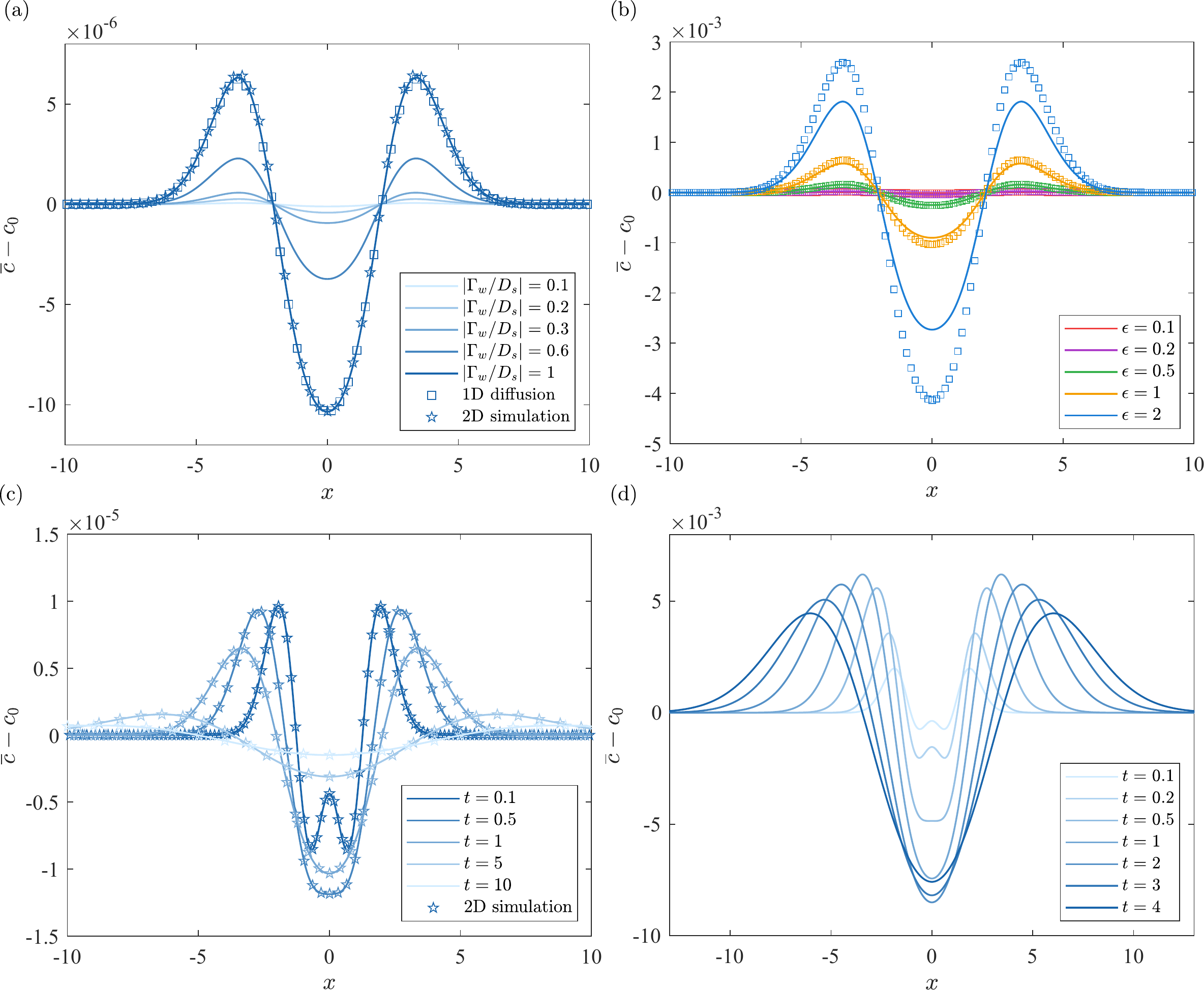}
\caption{Evolution of the cross-sectionally averaged solute dynamics for the 2D channel flow case. Results show the cross-sectionally averaged solute concentration $\overline{c}$ minus the results from pure diffusion $c_0$. (a) Results as a function of $|\Gamma_w/D_s|$ with $\epsilon=0.1$ and $t=1$. Solid lines correspond to the cross-sectionally averaged theoretical results developed in \S\ref{sec:ave}. Square symbols correspond to the numerical solution of the 1D forced diffusion equation given by (\ref{1dforce}). Star symbols indicate the cross-sectional average of the full 2D numerical simulations. All three methods of calculating $\overline{c}-c_0$ show excellent agreement at small $\epsilon$. (b) Results as a function of $\epsilon$ with $|\Gamma_w/D_s|=1$ and $t=1$. Solid lines correspond to numerical results, and square markers indicate theoretical predictions. The errors in the theoretical predictions manifest graphically for $\epsilon\geq2$. (c) Results over time with fixed $\Gamma_w/D_s=-1$ and $\epsilon=0.1$. Solid lines correspond to the theoretical predictions, and star symbols indicate the cross-sectional average of the full 2D numerical simulations. (d) Numerical results of the 1D model over time with fixed $\Gamma_w/D_s=-1$ and $\epsilon=10$.}
\label{fig:ave}
\end{figure}

Finally, following the strategy of Taylor and Aris, we consider the dynamics of the cross-sectionally averaged concentration profile.  In $\S$\ref{sec:ave}, we derived a cross-sectionally averaged concentration equation that can be used to model the net effects of diffusioosmotic dispersion. These equations are relatively simpler 1D diffusion equations with an effective diffusion coefficient that depends on the channel geometry and captures the effects of the dispersion. Unlike in the relatively simpler case of pure Taylor dispersion, here, the effective diffusivity is no longer a constant but rather a function of both axial position in the channel and time, as the net effects of the dispersion evolve with time and space. Thus, several methods are available for characterizing the cross-sectionally averaged solute dynamics. First, this 1D variable diffusivity model can be easily numerically integrated in time to yield the dynamics. Second, the results of the 2D numerical simulations can be averaged over the cross-section. Finally, the theoretical results for $c=c_0+\epsilon^2 c_1$ can be averaged over the cross-section to yield the theoretical leading-order dynamics. All of these approaches should be expected to match up to $\mathcal{O}(\epsilon^2)$. A visualization of the cross-sectionally averaged solute dynamics is shown in figure \ref{fig:ave} for the parallel-plate channel. Here, the results are the deviation of the cross-sectionally averaged solute concentration from the 1D results i.e., $\overline{c}-c_0$. Figure \ref{fig:ave}(a) shows results as a function of $|\Gamma_w/D_s|$ with $\epsilon=0.1$ and $t=1$. The solid lines indicate the theoretical predictions. The square markers correspond to the results of numerically solving the 1D forced diffusion equation given by (\ref{1dforce}). The star markers represent the results of averaging the 2D numerical simulation results over the cross-section. All three methods show close agreement. Recall that in the effective diffusivity coefficients derived in $\S$\ref{sec:ave}, the contribution from diffusioosmosis is $\mathcal{O}(\Gamma_w/D_s)^2$, such that the sign of the mobility does not affect the cross-sectionally averaged evolution.

However, increasing the magnitude of $\Gamma_w/D_s$ enhances the diffusioosmosis relative to the solute diffusion, leading to greater dispersion and greater deviations from $c_0$. Figure \ref{fig:ave}(b) shows the cross-sectionally averaged solute concentration as a function of $\epsilon$ with $|\Gamma_w/D_s|=1$ and $t=1$. The solid lines indicate the numerical simulations, and the square markers correspond to the theoretical predictions. As the value of $\epsilon$ increases, the magnitude of diffusioosmotic dispersion increases, enhancing the deviations from the pure diffusion case. The theoretical predictions show a good agreement with the simulations up to $\epsilon=2$. Figures \ref{fig:ave}(c,d) present the cross-sectionally averaged concentration profile as a function of time. Figure \ref{fig:ave}(c) shows the theoretical (solid lines) and numerical (star symbols) predictions of the cross-sectionally averaged solute concentration for constant $\Gamma_w/D_s=-1$ and $\epsilon=0.1$ and (d) shows the cross-sectionally averaged solute concentration from numerical simulations for constant $\Gamma_w/D_s=-1$ and $\epsilon=10$. As time progresses, a relative depletion zone forms near the channel center that is balanced by accumulation regions to the left and right of the solute peak. Note that the depletion at $x=0$ does not form quite as quickly as that near $x=0$, resulting in a small bump that decays with time. This is due to the fact that the axial concentration gradient is zero at $x=0$ due to symmetry, such that the wall slip velocity and any recirculating flow is negligible very near the centerplane. Thus, some time is still required for the solute to diffuse and adjust.

\begin{figure}
\centering
\includegraphics[width=\textwidth]{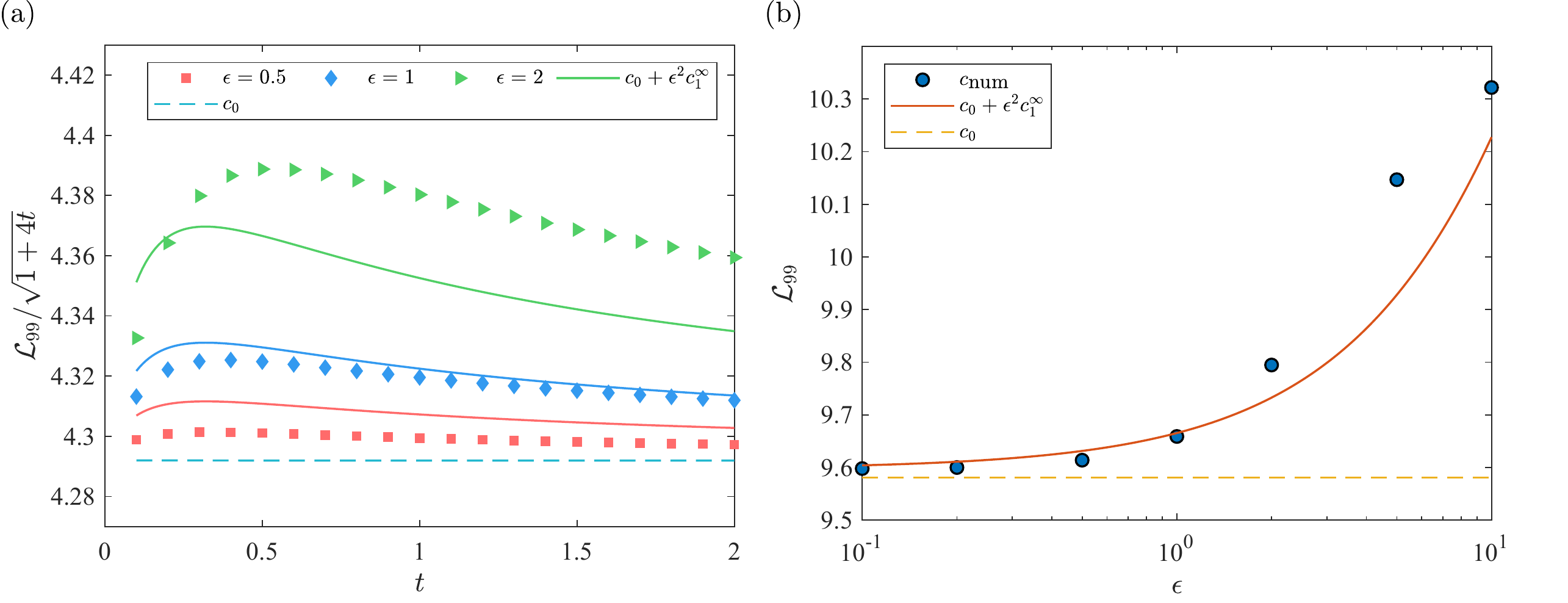}
\caption{Rescaled width of the cross-sectionally averaged higher-order solute concentration $\mathcal{L}_{99}/\sqrt{1+4t}$ with $\Gamma_w/D_s=-1$. The distribution width is defined such that $\bar c(\mathcal{L}_{99},t)=0.01\,\bar c(0,t)$. (a) Transient evolution of the distribution width as a function of $\epsilon$. Symbols correspond to numerical simulation results, and the solid and light blue dashed lines correspond to the theoretical predictions of the spread of $c0+\epsilon^2c_1^\infty$ and $c_0$, respectively. (b) Distribution width at $t=1$ as a function of $\epsilon$. The results show that the theory works well when $\epsilon<1$. For increasing $\epsilon$, the diffusioosmosis enhances the rate of spreading of the solute pulse, but this effect decays over time as the solute gradient weakens.}
\label{fig:distribution_width}
\end{figure}

In addition, we study the width of the spread of the cross-sectionally averaged concentration profiles. Here, we define the width of the solute distribution as $\mathcal{L}_{99}$, which corresponds to the axial position ($x$ or $z$), where the concentration has decayed by 99\% of its current peak value. Numerical results for a range of $\epsilon$ values are shown in figure \ref{fig:distribution_width} and compared to the theoretical predictions. Panel (a) shows the rescaled width of the concentration distributions as functions of time with constant $\Gamma_w/D_s=-1$ for different values of $\epsilon$. The colored markers represent the numerical results of solute concentration, and the solid and light blue dashed lines correspond to theoretical predictions of the distribution width of $c0+\epsilon^2c_1^\infty$ and $c_0$, respectively. Here, the distribution widths are rescaled by $\sqrt{1+4t}$, which represents the expected spreading rate of $c_0$. As can be seen, the results tend toward constant values, indicating that the spread of the higher-order solute profile is set by the spread of $c_0$. This is due to the fact that while $c_0$ becomes small at the edge of the distribution, the gradient of the logarithm of $c_0$ may remain large, allowing the diffusioosmotic dispersion to act over a relatively wider distribution. As $\epsilon$ increases, the width of the higher-order solute distribution increases. Figure \ref{fig:distribution_width}(b) shows $\mathcal{L}_{99}$ as a function of $\epsilon$ with constant $\Gamma_w/D_s=-1$ at a fixed time of $t=1$. Here, the theoretical predictions of $c_0+\epsilon^2c_1^\infty$ and $c_0$ are shown as a solid line and a dashed line, respectively. The symbol markers correspond to the numerical results. At a given time, the width of the distribution increases with a larger $\epsilon$ value. The theoretical predictions show a good agreement with numerical simulation for $\epsilon<1$, and the error between simulations and theory for this metric remains less than 10\% even at $\epsilon = 10$.

\section{Conclusion}\label{sec:Discussion and Conclusion}
In this study, we investigated the diffusioosmotic dispersion in a long, narrow channel with an initial Gaussian plug of solute at the center of the channel. The concentration gradients associated with the diffusion of this solute induce diffusioosmotic slip flow at the channel walls. The slip flow, in turn, drives recirculation inside the channel so that the transport of the solute is not purely diffusive, but also advective. The recirculating fluid flow in the channel introduces a shear flow that distorts the solute concentration profile and alters the effective transport of it analogously to the process of Taylor dispersion. We derived theoretical solutions for the coupled fluid and solute dynamics in such systems for both 2D channel flows and axisymmetric pipe flows. The theoretical derivation utilized a multiple-timescale analysis that incorporates both the fast and slow time dynamics. By averaging the governing equations over the cross section and incorporating the leading-order solutions, the system was reduced to a 1D diffusion equation with a variable effective diffusivity coefficient that depends on the geometry of the system and incorporates the effects of diffusioosmotic dispersion in an averaged sense. This approach is analogous to that of Taylor and Aris in the study of Taylor dispersion. Numerical solutions of this effective 1D model were compared with cross-sectionally averaged results of the 2D numerical simulations as well as the theoretical averaged results, all of which show good agreement. As much as possible, we have kept this analysis independent of the specific initial condition, except for the limitation that it be uniform over the cross section and have a characteristic distribution that is reasonably spread out relative to the channel width. The specific Gaussian initial condition chosen here was only needed for the calculation of $B(x,t)$ due to the Fourier transform approach needed to solve for it. Thus, while the full derivation of the effective diffusivity coefficients is not completely general, it can easily be adapted to other systems and initial conditions by modifying this one section of the calculation. It may also be possible to seek an explicit solution to $B(x,t)$ that remains general of the initial solute concentration, but this is a problem that we leave for future work. The results and analysis presented here have documented the role of diffusioosmosis-driven recirculation on the transport of solute in a straight channel or pipe flow and have shown how the cross-sectionally averaged effective dynamics can be reduced to a 1D effective diffusion problem analogously to Taylor dispersion.

\appendix

\section{Numerical details}\label{Appendix:A}
\begin{figure}
    \centering
    \includegraphics[width=\textwidth]{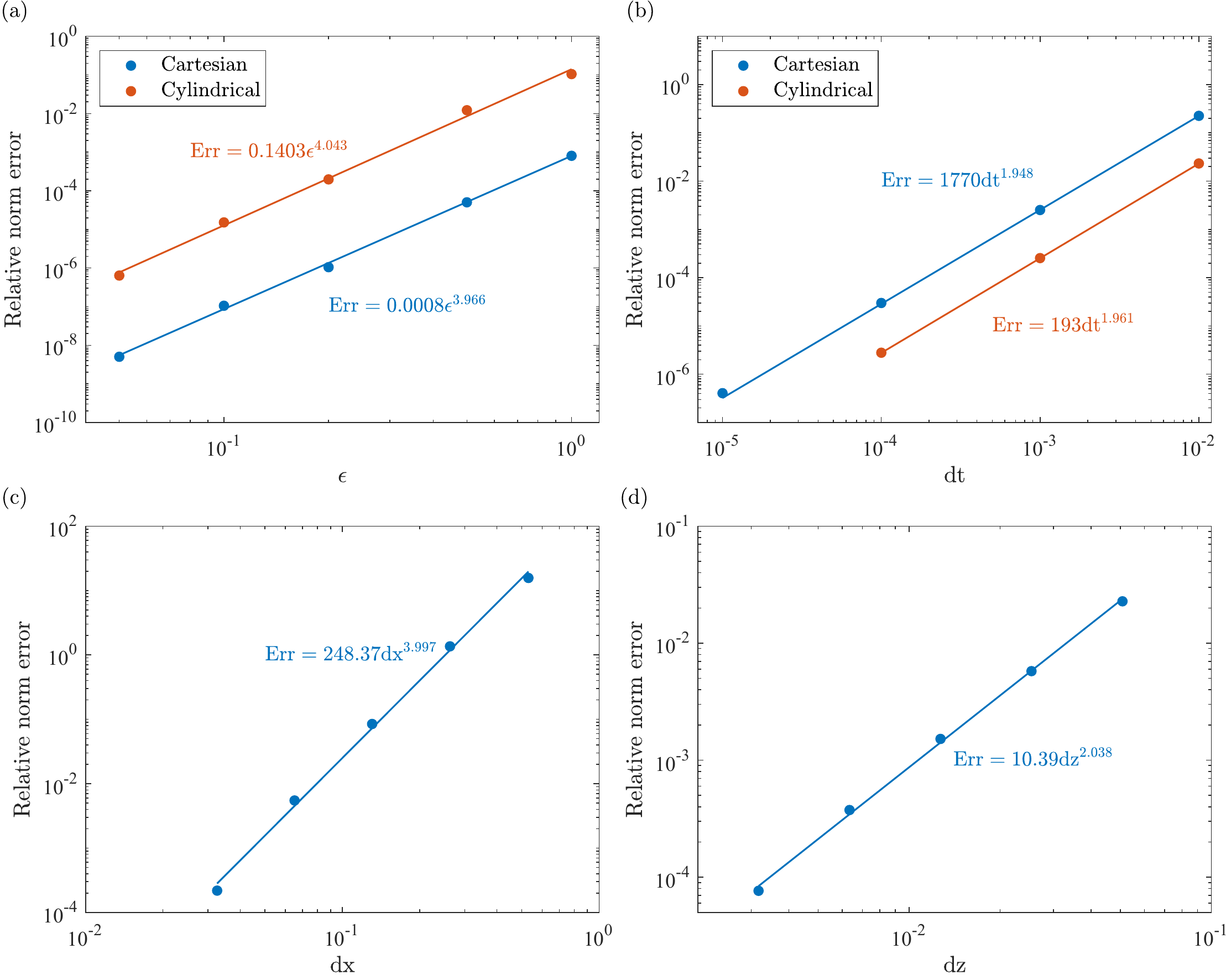}
    \caption{The relative norm error in cylindrical and Cartesian convergence studies with $\Gamma_w/D_s = -1$. The solid lines are the best-fit power law curves, and the matching color equations are the corresponding fitted power law functions. (a) Relative norm error between the theoretical and numerical prediction of $c_0+\epsilon^2c_1$ as a function of $\epsilon$ at $t=0.01$. (b) Convergence test results with respect to the time step $dt$ with grid size as $2049\times1025$ for the cylindrical case and $1600\times200$ for the Cartesian case. Here, the final time is 0.1 and $\epsilon=0.1$. (c) Spatial convergence test results for the Cartesian case with respect to $dx$. (d) Spatial convergence tests for the cylindrical case with respect to $dz$. Here, we used $\epsilon=0.1$, $\textrm{dt} = 1\times10^{-5}$ and $t_\textrm{final} = 0.1$ for spatial convergence studies.}
    \label{fig:convergence}
\end{figure}

In this section, we will discuss the numerical convergence tests. In the numerical simulations for the Cartesian case, we implemented a numerical scheme that is second-order accurate in time and fourth-order accurate in space. For the cylindrical case, we implemented a second-order accurate scheme in both time and space. Figure \ref{fig:convergence} shows the numerical convergence study results with $\Gamma_w/D_s = -1$. The solid lines are the best-fit power law curves, and the matching color equations are the corresponding fitted power law functions. Figure \ref{fig:convergence}$a$ shows the relative norm error between the theoretical and numerical prediction of $c_0+\epsilon^2c_1$ at $t=0.01$ as we increase the value of $\epsilon$. The theoretical prediction of solute concentration $c_0+\epsilon^2c_1$ has correction terms due to diffusioosmosis, which are order $\epsilon^4$. The theoretical predictions converge to $c_0$ as $\epsilon$ goes to zero, which can also be observed in figure \ref{fig:convergence}$a$. Figure \ref{fig:convergence}$b$ shows the order of convergence in the time step, $dt$, for both Cartesian and cylindrical cases with $\epsilon=0.1$. Here, we fixed the cylindrical case grid size as $2049\times1025$ and the Cartesian case grid size as $1600\times200$. In both cases, the simulation ran until $t=0.1$. The relative norm error is calculated in reference to the smallest $dt$ case in the simulation. Both Cartesian and cylindrical cases are second-order accurate in time as shown in the best-fit curve in figure \ref{fig:convergence}$b$, where the absolute norm error is shown to decrease as $O(\textrm{dt}^2)$, as expected. We chose $\textrm{dt}=10^{-4}$ for both cases as the relative error is less than $10^{-4}$. Figure \ref{fig:convergence}$c$ and \ref{fig:convergence}$d$ are the spatial convergence studies with $\epsilon=0.1$ for the Cartesian and cylindrical cases, respectively. Here, we used $\textrm{dt} = 1\times10^{-5}$ and $t_\textrm{final} = 0.1$ for both Cartesian and cylindrical spatial convergence studies. The relative norm error is calculated in reference to the finest grid in the simulation. As shown in figures \ref{fig:convergence}$c$ and \ref{fig:convergence}$d$, the errors are $O(\textrm{dx}^4)$ in the Cartesian case and $O(\textrm{dz}^2)$ in the cylindrical case.

\section{Derivation of cylindrical channel}\label{Appendix:B}
In this section, we demonstrate the theoretical solution to the fluid and solute dynamics in the axisymmetric cylindrical coordinate system. The general procedure is the same as that for the 2D Cartesian coordinate system. The governing equations for the system are provided by (\ref{dim_governing}). Here, we consider an axisymmetric configuration. We introduce the following nondimensionalizations:
\begin{equation}\label{eq:non_dim_co_cyl}
    z = \frac{z^*}{\ell},\; r = \frac{r^*}{a},\; u_z = \frac{u_z^*}{U},\; u_r = \frac{u_r^*\ell}{Ua},\; P = \frac{P^* a^2}{\mu U \ell},\; \epsilon = \frac{a}{\ell},\;\; U = \frac{D_s}{\ell},\; t = \frac{t^*}{\ell^2/D_s}.
\end{equation}
With these scalings the nondimensional form of the governing equations in cylindrical coordinates are given by
\begin{subequations}\label{eq:cylindrical_goven}
\begin{gather}
    \frac{1}{r}\frac{\partial (ru_r)}{\partial r} + \frac{\partial u_z}{\partial z} = 0,\\
    -\frac{\partial p}{\partial r} + \epsilon^2\frac{1}{r}\frac{\partial}{\partial r}\left(r\frac{\partial u_r}{\partial r}\right) -\epsilon^2\frac{u_r}{r^{2}}+ \epsilon^4\frac{\partial^2u_r}{\partial z^{2}} = 0,\\
    -\frac{\partial p}{\partial z} + \frac{1}{r}\frac{\partial}{\partial r}\left(r\frac{\partial u_z}{\partial r}\right) + \epsilon^2\frac{\partial^2u_z}{\partial z^{2}}=0,\\
    \epsilon^2\frac{\partial c}{\partial t} + \epsilon^2 u_r \frac{\partial c}{\partial r} + \epsilon^2 u_z \frac{\partial c}{\partial z} = \frac{1}{r}\frac{\partial}{\partial r}\left(r\frac{\partial c}{\partial r}\right) + \epsilon^2\frac{\partial^2 c}{\partial z^{2}}.
\end{gather}
\end{subequations}
The solution to the governing equation (\ref{eq:cylindrical_goven}) is subject to boundary conditions on the fluid and solute. These boundary conditions can be summarized by:
\begin{equation}
    \text{Quiescent far-field conditions:}~p=0~~\text{and}~~~u_z = 0~~~\text{and}~~~c=0~~~\text{at}~z=\pm\infty,
\end{equation}
\begin{equation} \label{bc:v_cyl}
    \text{No fluid penetration at the walls: } u_r = 0~\text{at}~r=1,
\end{equation}
\begin{equation} \label{c_bc_cyl}
    \text{No-flux conditions at the channel walls: }\frac{\partial c}{\partial r}=0~\text{at}~r=1,
\end{equation}
\begin{equation} \label{slip_condition_cyl}
\text{Diffusioosmotic wall slip boundary condition: }u_z = \frac{\Gamma_w}{D_s}\frac{\partial \ln c}{\partial z}~\text{at}~r=1.
\end{equation}

Similar to the Cartesian case, we introduce a multiple timescale approach \citep{bender1999advanced} in which we introduce a fast time variable $T = t/\epsilon^2$. That is, $T=O(1)$ over dimensional times $\sim a^2/D_s$ corresponding to $t=O(\epsilon^2)$, whereas $t=O(1)$ on dimensional times $~\ell^2/D_s$. Thus, we rewrite the advection-diffusion equation as,
\begin{equation}\label{adv_diff_cyl}
    \epsilon^2\frac{\partial c}{\partial t} + \frac{\partial c}{\partial T}+ \epsilon^2 u_r \frac{\partial c}{\partial r} + \epsilon^2 u_z \frac{\partial c}{\partial z} = \frac{1}{r}\frac{\partial}{\partial r}\left(r\frac{\partial c}{\partial r}\right) + \epsilon^2\frac{\partial^2 c}{\partial z^{2}}.
\end{equation}
Again, similar to the approach used in Cartesian coordinates, we seek the analytical solution of the governing equations as perturbation expansions with the small parameter $\epsilon = a/\ell$ using an expansion that has the form
\begin{subequations}\label{perturbation_cyl}
\begin{align}
c(r,z,t,T) &= c_0(r,z,t,T) + \epsilon^2 c_1(r,z,t,T) + \epsilon^4 c_2(r,z,t,T) + \hdots,\\
p(r,z,t,T) &= p_0(r,z,t,T) + \epsilon^2 p_1(r,z,t,T) + \epsilon^4 p_2(r,z,t,T) + \hdots,\\
u_z(r,z,t,T) &= u_{z0}(r,z,t,T) + \epsilon^2 u_{z1}(r,z,t,T) + \epsilon^4 u_{z2}(r,z,t,T) + \hdots,\\
u_r(r,z,t,T)  &= u_{r0}(r,z,t,T) + \epsilon^2 u_{r1}(r,z,t,T) + \epsilon^4 u_{r2}(r,z,t,T) +\hdots.
\end{align}
\end{subequations}
We first need to obtain the leading-order velocity and pressure solutions. These can be obtained by substituting equations (\ref{perturbation_cyl}) into the governing equations, which gives
\begin{subequations}\label{leading_order_cyl}
\begin{gather}
    \frac{\partial u_{z0}}{\partial z} + \frac{u_{r0}}{r} + \frac{\partial u_{r0}}{\partial r} = 0,\\
    \frac{\partial p_0}{\partial r} = 0,\\
    -\frac{\partial p_0}{\partial z} + \frac{1}{r}\frac{\partial u_{z0}}{\partial r} + \frac{\partial^2 u_{z0}}{\partial r^2} = 0,\\
    \frac{\partial c_0}{\partial T}=\frac{1}{r}\frac{\partial}{\partial r}\left(r\frac{\partial c_0}{\partial r}\right),
\end{gather}
\end{subequations}
to leading order. Considering the fact that the initial condition is given by $c_0(T=0,t=0)=\exp(-z^2)$ and that there are no-flux conditions at the channel walls, i.e., $\frac{\partial c_0}{\partial r}(r=\pm 1)=0$, it must be true that $c_0(r,z,t,T)=c_0(z,t)$, with $c_0(t=0)=\exp(-x^2)$.

To solve for the leading-order velocities, we first take equation (\ref{leading_order_cyl}$a$) and solve for $u_{z0}$, which is subject to the slip boundary condition (\ref{slip_condition_cyl}). The leading-order $u_{z0}$ becomes
\begin{equation} \label{uz}
    u_{z0} = \frac{\Gamma_w}{D_s c_0}\frac{\partial c_0}{\partial z} + \frac{1}{4}(-1+r^2)\frac{\partial p_0}{\partial z}.
\end{equation}
Here, $u_{z0}$ has a term that includes $p_0(z,t)$, which can be found by considering the conservation of mass and integrating over the channel cross-section,
\begin{equation}
    \int_0^1 u_{z0}2\pi r \,dr = 0.
\end{equation}
Following this approach, $p_0(z,t)$, is found to be
\begin{equation}\label{p0_cyl}
    p_0 = 8\frac{\Gamma_w}{D_s}\ln(c_0).
\end{equation}
With equation (\ref{p0_cyl}), $u_{z0}(r,z,t)$ can be simplified and written as
\begin{equation}\label{eq:uz0}
    u_{z0} = \frac{\Gamma_w}{D_s c_0}(-1+2r^2)\frac{\partial c_0}{\partial z}.
\end{equation}
To solve for $u_{r0}(r,z,t)$, we integrate the continuity equation (\ref{leading_order_cyl}a) and apply the boundary condition given by (\ref{bc:v}). This gives the leading-order $u_{r0}$ to be
\begin{equation}\label{eq:ur0}
    u_{r0} = \frac{\Gamma_w r(-1+r^2)}{D_s2c_0^2}\left(\left(\frac{\partial c_0}{\partial z}\right)^2-c_0\frac{\partial^2 c_0}{\partial z^2}\right).
\end{equation}
Substituting the asymptotic expansion into the advection-diffusion equation, we find, to leading-order, that
\begin{equation}\label{leading_adv_cyl}
     \frac{\partial c_0}{\partial t}+\frac{\partial c_1}{\partial T}+u_{z0}\frac{\partial c_0}{\partial z}-\frac{\partial^2 c_0}{\partial z^2}- \frac{1}{r}\frac{\partial c_1}{\partial r} - \frac{\partial^2 c_1}{\partial r^2} = 0,
\end{equation}
subject to the initial condition $c_0(t=0)=\exp(-z^2)$ and $c_1(t=T=0)=0$. To solve this, we note that at long times the fast-time dynamics should have all decayed such that the time derivative term can be ignored, and the equation can be averaged across the channel to obtain
\begin{equation}
    \frac{\partial c_0}{\partial t} = \frac{\partial^2 c_0}{\partial z^2}.
\end{equation}
As in the Cartesian case, the solution to this is
\begin{equation} \label{initial_cyl}
    c_0(z,t)=\frac{1}{\sqrt{1+4t}}\exp{\left(-\frac{z^2}{1+4t}\right)}.\\
\end{equation}
Substituting this into (\ref{leading_adv_cyl}) gives
\begin{equation}
     \frac{\partial c_0}{\partial t}+\frac{\Gamma_w}{D_s c_0}(-1+2r^2)\left(\frac{\partial c_0}{\partial z}\right)^2+\frac{\partial c_1}{\partial T}-\frac{\partial^2 c_0}{\partial z^2}- \frac{1}{r}\frac{\partial c_1}{\partial r} - \frac{\partial^2 c_1}{\partial r^2} = 0.
\end{equation}
At long times, the fast time dynamics have decayed such that derivatives with respect to $T$ can be neglected, and the equation can be integrated to yield
\begin{equation}\label{c1infty_cyl}
    c_1(T\rightarrow\infty)\sim c_1^\infty(r,z,t) = \frac{1}{8}r^2\left[2\frac{\partial c_0}{\partial t} + \frac{\Gamma_w}{D_s c_0}(-2+r^2)\left(\frac{\partial c_0}{\partial z}\right)^2 - 2\frac{\partial^2 c_0}{\partial z^2}   \right] + B(z,t),
\end{equation}
where $B(z,t)$ is a yet unknown function obtained from integration. This can be determined by applying conservation of mass and taking the cross-sectional average of the advection-diffusion equation, which gives
\begin{equation}
    \frac{\partial B}{\partial t} = \frac{\partial^2 B}{\partial z^2}-\frac{e^{\frac{-z^2}{1+4t}}\Gamma_w/D_s(2+32t^2+4t(4+\Gamma_w/D_s z^2)+\Gamma_w/D_s(z^2-z^4))}{3(1+4t)^{9/2}}.
\end{equation}
Using a Fourier transform approach, the solution can be found to be
\begin{equation}
    B = \frac{2\Gamma_w/D_s\left[16z\alpha^2+4\Gamma_w/D_s t(3\alpha^2-12\alpha z^2+4z^4)-3\Gamma_w/D_s\alpha^2(\alpha-2z^2)\ln\alpha \right]}{e^{\frac{z^2}{\alpha}}(\alpha^{7/2}(96+384t))}.
\end{equation}
This equation does not satisfy the initial condition, so it is not yet the full solution for the higher-order solute dynamics. Now, we look for a solution $c_1(r,z,t,T) = c_1^\infty(r,z,t)+\hat{c}_1(r,z,t,T)$. Substituting into equation (\ref{adv_diff_cyl}), we find
\begin{equation}
    \frac{\partial \hat{c_1}}{\partial T} = \frac{1}{r}\frac{\partial}{\partial r}\left(r\frac{\partial \hat{c}_1}{\partial r}\right),
\end{equation}
with the initial condition given by
\begin{equation}
    \hat{c}_1(T=0) = -c_1^\infty(t=0)=-\frac{1}{6}e^{-z^2}\frac{\Gamma_w}{D_s}(2-6r^2+3r^4)z^2.
\end{equation}
The solution to this is in the form of a Bessel series $\hat{c}_1(r,T) = \sum_{n=0}^\infty a_n e^{-\lambda_n^2T}J_0(\lambda_n r)$, with initial condition of $f(r) = -\frac{1}{6}e^{-z^2}\frac{\Gamma_w}{D_s}(2-6r^2+3r^4)z^2$ and boundary condition of $\frac{\partial \hat{c_1}}{\partial r}=0$ at $r=1$. Using the boundary condition, the $\lambda_n$ can be found to be the roots of $J_1$, which we denote as $j_{1,n}$. The coefficients $a_n$ can be determined as follows
\begin{equation}
\begin{split}
    a_n &= \frac{2}{J_0(j_{1,n})^2}\int_0^1 r f(r)J_0(j_{1,n} r) \,dr\\
    &= -\frac{e^{-z^2}\frac{\Gamma_w}{D_s}z^2(96J_2(j_{1,n})-J_1(j_{1,n})j_{1,n}(24+j_{1,n}^2)) }{3J_0(j_{1,n})^2j_{1,n}^4}.
\end{split}
\end{equation}
Thus, the solution of $\hat{c}_1$ is as follows
\begin{equation}
    \hat{c}_1(r,z,t,T) = 2\int_0^1 rf(r)\,dr+\sum_{n=1}^\infty a_n e^{-j_{1,n}^2 T}J_0(j_{1,n}r).
\end{equation}
We can then construct a composite solution $c_1 = c_1^\infty+\hat{c}_1$ that is valid for all $t$ using the fact that $T = t/\epsilon^2$.

\bibliography{references}
\end{document}